\documentclass[aps,prd,multicol,groupedaddress,nofootinbib,preprint]{revtex4}
\usepackage{amsmath}
\usepackage[dvips]{graphicx}

\begin{document}
\preprint{YITP-05-24}
\date{\today}
\title{Perturbative Study of the Supersymmetric Lattice Theory
from Matrix Model}
\author{Tetsuya Onogi}
\email[e-mail: ]{onogi@yukawa.kyoto-u.ac.jp}
\author{Tomohisa Takimi}
\email[e-mail: ]{takimi@yukawa.kyoto-u.ac.jp}
\affiliation{Yukawa Institute for Theoretical Physics, Kyoto University,
Kitashirakawa-Oiwakecho, Sakyo, Kyoto 606-8502, Japan}
\numberwithin{equation}{section}
\begin{abstract}
We study the lattice model for the supersymmetric Yang-Mills theory 
in two-dimensions proposed by Cohen, Kaplan, Katz, and Unsal. 
We re-examine the formal proof for the absence of susy breaking counter 
terms as well as the stability of the vacuum by an explicit 
perturbative calculation for the case of $U(2)$ gauge group.
Introducing fermion masses and treating the bosonic zero momentum mode 
non-perturbatively, we avoid the infra-red divergences in the
perturbative calculation. As a result, we find that there 
appear mass counter terms for finite volume which vanish
in the infinite volume limit so that the theory needs no fine-tuning. 
We also find that the supersymmetry plays an important role in 
stabilizing the lattice spacetime by the deconstruction. 

\end{abstract}
\maketitle

\section{Introduction}

The lattice field theory methods are expected to be useful 
for the non-perturbative study of supersymmetric gauge theories,
but a satisfactory formulation which can be applied to efficient 
simulation has not been obtained so far despite much effort in the study 
of supersymmetric lattice formulations
(see 
~\cite{Kaplan2}-\cite{Neuberger 98}
). 
Since the supersymmetry algebra contains infinitesimal translations, 
it is quite difficult to construct a lattice theory without 
an explicit breaking of the supersymmetry due to the lattice 
regularization, which in principle gives rise to all 
possible supersymmetry breaking terms not 
prohibited by other symmetries at the quantum level. 
This makes practical non-perturbative simulations extremely difficult
due to too many parameters which requires fine-tuning in order to 
recover the supersymmetry in the continuum limit. 

To solve this problem, one of the promising approaches is to construct the 
lattice formulations preserving partial exact supersymmetry
\footnote {The first attempt to 
construct a theory with partial exact supersymmetry on the lattice 
was proposed by Sakai-Sakamoto~\cite{Sakai 83}. In recent years, 
not only CKKU model but also other
several lattice formulations for Yang-Mills theories with an exact partial 
supersymmetry have been proposed. One approach is the topological field 
theory (TFT) construction of the lattice theory, which can be obtained 
from twisting the gauge theory with extended 
supersymmetry~\cite{Catterall:2003wd,Catterall:2004oc}, 
\cite{Sugino1}-\cite{Sugino4}.} .
Cohen-Kaplan-Katz-Unsal(CKKU)~\cite{Kaplan2,Kaplan1,Kaplan3,Kaplan:2005ta}
constructed a matrix model realization of such theories based on the 
\textit{orbifolding}~\cite{Orbifold1,Orbifold2} and 
\textit{deconstruction}~\cite{deconstruction1,deconstruction2} 
method. 
The first non-perturbative studies of these CKKU models are performed by 
Giedt~\cite{Giedt 0304}
-\cite{Giedt 0405}.
\textit{Orbifolding} is the projection of the 
zero-dimensional 
or one-dimensional 
matrix models by some discrete subgroup of the symmetry. 
A zero-dimensional moose diagram which is regarded as lattice structure is 
obtained by this procedure. 
Supercharges which are invariant under the orbifold projection becomes 
the symmetry on the lattice. In these procedures, one can make lattice 
models with an exact partial supersymmetry 
if one chooses appropriate generators 
for orbifold projection. 
\textit{Deconstruction} is a dynamical 
construction of the $d$-dimensional spacetime on a $N^d$ lattice with the 
spontaneous symmetry breakdown of the gauge symmetry of the moose diagram 
$U(M)^{N^d}\rightarrow U(M)_{diag}$. In this model, they apply the 
deconstruction at the zero-dimensional moose diagram. 

One possible problem in this approach is that the extended 
supersymmetry has flat directions for the scalar so that 
the lattice structure from the deconstruction suffers from 
the instability due to the quantum fluctuations of the scalar zero momentum 
modes. 
To suppress the divergence in the flat directions, 
soft susy-breaking terms for the scalar fields are introduced. 
Since such terms 
break the supersymmetry 
and causes the infra-red divergence 
of fermion zero modes, 
the original discussion of the renormalization based on 
exact supersymmetry on the lattice has to be modified 
by including the breaking terms. 

In this paper, we concentrate on the 
two-dimensional 
$U(2)$ lattice gauge 
model of CKKU in Ref.~\cite{Kaplan2},
and investigate the fine-tuning problem and the stability of the 
spacetime structure by an explicit calculation of quantum corrections 
of fields which can be relevant. We calculate the quantum corrections 
of scalar one-point and two-point functions in the model of 
Ref.~\cite{Kaplan2}. 
Before the explicit calculation, 
we have to take care of ill-defined perturbation 
due to the flat directions in the zero momentum modes of gauge fields 
and fermion fields~\cite{Giedt 0304}.
In order to avoid the infra-red 
divergence for the fermion zero mode, we introduce a new soft susy 
breaking mass term for the fermion fields. For the bosonic fields, 
we apply the perturbation only for the non-zero momentum mode and treat 
the zero momentum mode non-perturbatively. 
In addition to the fine-tuning problem, 
several interesting results
are obtained by our explicit calculation.
Firstly, 
we found the constraint for the parameter region 
where the lattice 
theory is well-defined. 
And secondly, it is found 
that the fermion-boson cancellation 
which suppresses the quantum corrections to the potential 
is needed 
to stabilize the deconstructed spacetime 
in the physical region where the lattice size is larger than the 
correlation length.
Similar instability has been observed in the non-perturbative study 
~\cite{Giedt 0312} on the
bosonic part of the CKKU model for the (4,4) 2d
super-Yang-Mills~\cite{Kaplan3}.

The paper is organized as follows.
We review the model by CKKU \cite{Kaplan2} in Sec.~\ref{Sec:review}. 
In Sec.~\ref{Sec:method}, we explain possible 
counter terms.
We also explain the problem of fermion zero-mode which is called as  
^^ ever-existing fermion zero mode'.
In Sec.~\ref{Sec:method2}, 
we will describe the treatment of massless zero momentum modes 
which make the perturbative calculation
based on the gaussian integral
ill-defined. 
In Sec.~\ref{Sec:non0}, 
we present our results on 
the renormalization of susy breaking counter terms.
Sec.~\ref{decon} is devoted to the 
discussion on the constraint from the stability of the spacetime. 
Our conclusion and discussions are given in Sec.~\ref{Sec:summary}. 
Technical details such as mathematical notations, 
path-integral measures, and amplitudes are 
described in the Appendices.

\section{Brief review of CKKU model}\label{Sec:review}
The model by CKKU~\cite{Kaplan2} is constructed from the zero-dimensional 
matrix model with four supercharges, 
\begin{equation}
S=\frac{1}{g^2}(\frac{1}{4}Trv_{mn}v_{mn}
+Tr\bar{\psi}\bar{\sigma}_m[v_m,\psi]),
\end{equation}
where $\sigma$ is the Pauli matrices, $v_{mn}=[v_m,v_n]$ 
and $v_m=v_m^{\alpha}T^{\alpha},\, \psi=\psi^{\alpha}T^{\alpha}$,
$T^{\alpha}$ 
is the generators of $U(MN^2)$ 
gauge group,
and $g$ is the gauge coupling. 
The above action is obtained by the dimensional 
reduction of the 4-dimensional $\mathcal{N}=1$ 
super-Yang-Mills theory to the zero-dimensional theory.
They constructed the lattice structure by imposing the following orbifolding 
condition on the matrix theory 
\begin{equation}
\Phi_{\mu,\nu}
=[e^{\frac{2\pi ir_a}{N}}\mathcal{C}_a\Phi \mathcal{C}_a^{-1}]_{\mu,\nu},
\end{equation}
where  $\mu,\nu$ are the 
indices of the gauge group $U(MN^2)$. 
$r_a(a=1,2)$ are the generators of the Cartan subalgebra of the R-symmetry 
$SO(4)\times U(1)$,
whereas 
 $\mathcal{C}_a$ are generators of a discrete symmetry 
$Z_N\otimes Z_N \subset U(MN^2)$ 
as given in Ref.~\cite{Kaplan2}.
If we decompose the matrices into $N^2\times N^2$ blocks 
of $M\times M$ submatrices, the above orbifolding conditions require 
that only $N^2$ blocks can be non-zero, while the other blocks are projected 
out. By interpreting the indices for different blocks as 
the coordinates of the 
two-dimensional spacetime, we obtain a lattice structure which preserve 
one of the four supercharges exactly.
In this interpretation $N$ 
is regarded as the number of lattice sites for each directions.
The lattice action is 
\begin{eqnarray}
&S_0=
\frac{1}{g^2}\sum_\mathbf{n} 
Tr[\frac{1}{2}
(\bar{x}_\mathbf{n-i} x_\mathbf{n-i}-x_\mathbf{n}\bar{x}_\mathbf{n}
+\bar{y}_\mathbf{n-j}y_\mathbf{n-j}
-y_\mathbf{n}\bar{y}_\mathbf{n})^2\nonumber\\
&+2|x_\mathbf{n} y_\mathbf{n+i}-y_\mathbf{n} x_\mathbf{n+j}|^2\nonumber\\
&+\sqrt{2}(\alpha_\mathbf{n}\bar{x}_{\mathbf{n}}\lambda_\mathbf{n}
-\alpha_\mathbf{n-i}\lambda_\mathbf{n}\bar{x}_\mathbf{n-i})
+\sqrt{2}(\beta_\mathbf{n}\bar{y}_\mathbf{n}\lambda_\mathbf{n}
-\beta_\mathbf{n-j}\lambda_\mathbf{n}\bar{y}_\mathbf{n-j})\nonumber\\
&-\sqrt{2}(\alpha_\mathbf{n}y_\mathbf{n+i}\xi_\mathbf{n}
-\alpha_\mathbf{n+j}\xi_\mathbf{n} y_\mathbf{n})
+\sqrt{2}(\beta_\mathbf{n}x_\mathbf{n+j}\xi_\mathbf{n}
-\beta_\mathbf{n+i}\xi_\mathbf{n} x_\mathbf{n})],
\label{action}
\end{eqnarray}
where $x_\mathbf{n}$, $\bar{x}_\mathbf{n}$ and 
$y_\mathbf{n}$, $\bar{y}_\mathbf{n}$  are the linear combinations 
of the submatrices in $v_1$, $v_3$ and  $v_2$, $v_4$ respectively.
$\alpha_\mathbf{n}$, $\beta_\mathbf{n}$ 
$\lambda_\mathbf{n}$, $\xi_\mathbf{n}$ 
are the submatrices in $\bar{\psi}$ and $\psi$ respectively.

A mechanism called as 
\textit{deconstruction} is applied in which the kinetic term is 
generated by a spontaneous breakdown of the gauge symmetry. 
The bosonic potential in the action after the orbifolding 
allows the following classical minimum as vacuum expectation 
values (VEV) $x_\mathbf{n}=y_\mathbf{n}=\bar{x}_\mathbf{n}
=\bar{y}_\mathbf{n}=\frac{1}{\sqrt{2}a}\times \mathbf{1}_k \label{vev}$
, where $a$ is the lattice spacing.
Expanding the bosonic fields around this VEV as
\begin{eqnarray}
x_\mathbf{n}
=\frac{1}{\sqrt{2}a}\times \mathbf{1}_k 
+\frac{s_{x\mathbf{n}}+iv_{x\mathbf{n}}}
{\sqrt{2}},
&&
\bar{x}_\mathbf{n}
=\frac{1}{\sqrt{2}a}\times \mathbf{1}_k 
+\frac{s_{x\mathbf{n}}-iv_{x\mathbf{n}}}
{\sqrt{2}},\nonumber\\
y_\mathbf{n}=
\frac{1}{\sqrt{2}a}\times \mathbf{1}_k +\frac{s_{y\mathbf{n}}+iv_{y\mathbf{n}}}
{\sqrt{2}},
&&
\bar{y}_\mathbf{n}
=\frac{1}{\sqrt{2}a}\times \mathbf{1}_k 
+\frac{s_{y\mathbf{n}}-iv_{y\mathbf{n}}}
{\sqrt{2}},\label{expansion}
\end{eqnarray}
the action acquires kinetic terms. Taking a naive $a\to 0$ limit 
the action
can be written as 
\begin{align}
S=&\frac{1}{g_2^2}\int d^2x Tr
\Bigl( |D_m s|^2+\bar{\psi}iD_m \gamma_m \psi 
+\frac{1}{4}v_{mn}v_{mn} \nonumber\\
&+i\sqrt{2}(\bar{\psi}_L [s, \psi_R] 
+\bar{\psi}_R [s^{\dagger},\psi_L]) +\frac{1}{2}[s^{\dagger},s]^2 \Bigr),
\end{align}
which is $\mathcal{N}=2$ $U(M)$ super-Yang-Mills theory in two-dimensions. 
In this paper, we concentrate on $U(2)$ gauge theories.
Here $g_2=ga$ is the two-dimensional gauge coupling and  
$s=\frac{s_x+is_y}{\sqrt{2}}$ and
$s^{\dagger}$ is hermitian conjugate of $s$. 
The definition of the fermion fields 
and gamma matrices are same as in Ref.~\cite{Kaplan2}.

In Ref.~\cite{Kaplan2}, the authors 
argued that 
the theory recovers the full supersymmetry without the need for 
fine-tuning. Let us here repeat their arguments.
The counter terms which can appear in the two-dimensional lattice theory have 
the following form
\begin{equation}
\delta S = \frac{1}{g_2^2} Tr \int d\theta \int d^2x \mathcal{C_O O}
\label{con1}
\end{equation}
The mass dimension of coupling $M(g_2)$ 
is $M(g_2)=1$. And $M(\int d^2x)=-2$,$M(\int d\theta)=\frac{1}{2}$.
If the operator $\mathcal{O}$ has dimension $M(\mathcal{O})=p$, 
mass dimension of coefficient $\mathcal{C_O}$ must be 
$M(\mathcal{C_O})=\frac{7}{2}-p$. 
In perturbation theory, the coefficient $\mathcal{C_O}$ 
can be expanded as 
\begin{equation}
\mathcal{C_O}=a^{p-\frac{7}{2}}\sum_l c_l(g_2^2a^2)^l,
\end{equation}
where  $l$ is the order of loop expansion 
and $c_l$ is the coefficient of $l$-th order. 
Therefore at $l$-loop, relevant operators must satisfy 
\begin{equation}
p \le \frac{7}{2}-2l
\label{dimension}
\end{equation}
At 1-loop level, 
only operators with 
dimensions $0 \le p \le \frac{3}{2}$ are relevant. 
Beyond 1-loop level, 
there is no relevant operator, since Eq.~(\ref{dimension}) 
allows only the negative mass dimensions. 
The operators which can satisfy this condition 
are only 1-point function of bosonic super-field $\mathbf{B}$, 
and 1-point 
function of fermionic one $\mathbf{F}$. 
$\mathbf{B}$ cannot give any contributions due to the 
Grassman parity,
$\int d\theta$. 
There are two candidates 
$\Lambda$ and $\Xi$ for the 
fermionic 1-point function, where 
$\Lambda$ and $\Xi$ are the superfields corresponding to 
$\lambda$ and $\xi$.
Since $\Xi$ is forbidden by $Z_2$ point symmetry, 
the only term which can be relevant is $\Lambda$, however 
we can ignore this term since it is the cosmological constant.  
As a result, there is no relevant operator \textit{due to 
supersymmetry and the discrete symmetry on the lattice.}
This naive power counting arguments give a formal proof 
for the emergence of the supersymmetry in the continuum limit 
without fine-tuning.
However, we should remark that the above argument assumes
that the perturbation 
theory is well-defined. 

The formalism of CKKU also assumes the symmetry breaking for the
deconstruction. However as they pointed out, 
the integral over the zero momentum modes of scalar fields is divergent, 
since there are flat directions in the action Eq.~(\ref{action}). 
This divergence causes a serious instability of the vacuum. 
In order to control the stability of the vacuum, they modified 
the theory and introduced soft scalar mass terms to suppress the divergence, 
\begin{equation}
S_1 = S_0 + \frac{a^2\mu^2}{g^2}\sum_{\mathbf{n}}Tr [(x_\mathbf{n}
\bar{x}_\mathbf{n}-\frac{1}{2a^2})^2
+(y_\mathbf{n}
\bar{y}_\mathbf{n}-\frac{1}{2a^2})^2],
\label{soft-term}
\end{equation}
where they take mass parameter $\mu$ to 
be inversely proportional to the lattice size 
$L\equiv Na$. 
Whether the above formal proof for renormalization 
remains valid even with the soft 
susy breaking term should be examined.
And also whether the perturbation is well-defined or not 
should be studied.

\section{Subtleties in CKKU model}
\label{Sec:method}
In this section, we consider the subtleties in CKKU theory.
In the discussion on the renormalization in the previous section, 
they assumed that the perturbation theory is well-defined. 
However after introducing the soft susy breaking terms, there appear 
infra-red divergences from massless fields which do not cancel 
with each other. 
It is therefore important to re-examine the renormalization at 
1-loop level by explicit calculations in order to see whether this 
theory really needs  fine-tuning or not.

Since there is no exact supersymmetry in the modified action,  we 
do not exploit the superfield formalism here, 
so that operators
$\mathcal{O}$ in this section do not contain the grassman coordinate 
$\theta$ any more as opposed to the operators $\mathcal{O}$ 
in the previous section. 
Radiative corrections induce the operator $\mathcal{O}$ of the 
following structure into the action
\begin{equation}
\delta S=\frac{1}{g_2^2}Tr \int d^2z \mathcal{C_O} \mathcal{O}.
\end{equation}
Relevant or marginal operators ($\mathcal{O}$) whose canonical dimension 
$M[\mathcal{O}]=p$ 
at the $l$-loop correction must satisfy 
\begin{equation}  
p \le 4-2l 
\end{equation}
At 1-loop level, relevant or 
marginal 
operators with dimensions $0\le p \le 2$ 
can arise.  
At 2-loop level, relevant operators with the dimension $p=0$ 
can arise. Beyond 2-loop, there is no relevant 
or marginal counter term. 
Since the operator with the dimension $p=0$ is the cosmological constant, 
it does not play any serious 
role in fine-tuning problems.  

Let us now focus on the 1-loop relevant or 
marginal counter-terms.
Since bosonic fields have dimension 1 and fermionic fields have dimension 
$\frac{3}{2}$, the candidates for such operators are 
bosonic 1-point and 2-point functions.
Although fermionic 1-point functions are possible from dimension counting, 
they 
are forbidden by Grassman parity.

Since 1-point functions of gauge fields 
are forbidden from Furry's theorem and the 
2-point ones are also forbidden by the gauge symmetry. 
Hence the only possible 
counter terms are 
\begin{itemize}
\item $<s_x>,<s_y>$ (scalar 1point functions),
\item $<s^2_x>,<s^2_y>$ (scalar 2point functions).
\end{itemize}

In what follows, we will discuss 
the renormalization 
of these two operators.

Another subtlety is the existence of an exact zero mode 
of the fermion matrix called 
^^ ever-existing zero mode'.
It was pointed out by 
Giedt~\cite{Giedt 0304}
that the constant mode of the $U(1)$ part of the fermion, 
which is independent of the bosonic field configurations, 
completely decouples from the theory. Therefore a naive path-integral 
of this model would be  ill-defined, unless one either removes this mode 
or introduce an infra-red regulator. The existence of this mode can be 
understood as follows: The fermionic part of the action for the mother theory 
is 
\begin{equation}
S_F = \frac{1}{g^2} Tr (\bar{\psi}\bar{\sigma}_m[v_m,\psi]), \label{m}
\end{equation}
where the fields are described by the adjoint representation of 
$U(2N^2)$ gauge group. It is obvious that the $U(1)$ component 
of $\psi$ is the exact zero mode. Since $\lambda$ in $\psi$ 
has a neutral charge for the R-symmetry $U(1)_{r_1}\times U(1)_{r_2}$,
The constant mode $Tr_{U(MN^2)} [ \psi^\alpha(T^\alpha) ] 
= \sum_{\mathbf{n}} Tr \lambda^0_{\mathbf{n}}$
survives as an exact zero mode in the daughter theory after orbifolding. 
In this work, to make path-integral well-defined,
we propose to introduce the following fermion mass term with coefficient 
$\mu_F$ proportional to $\frac{1}{L}$ so that the action now becomes
\begin{eqnarray}
S_2 & =& S_1 + \frac{a\mu_F \sqrt{2}}{g^2} Tr \sum_{\mathbf{n}} 
(\alpha_{\mathbf{n}} \bar{x}_{\mathbf{n}} \lambda_{\mathbf{n}}
+ \beta_{\mathbf{n}} \bar{y}_{\mathbf{n}} \lambda_{\mathbf{n}} 
-\alpha_{\mathbf{n}} y_{\mathbf{n+i}} \xi_{\mathbf{n}}
+ \beta_{\mathbf{n}} x_{\mathbf{n+j}} \xi_{\mathbf{n}}). 
\label{fermimass}
\end{eqnarray}

Note that this mass term (\ref{fermimass}) and the bosonic mass 
terms (\ref{soft-term}) play slightly different roles. 
The bosonic term 
(\ref{soft-term}) gives mass only to scalar fields but not to the gauge
fields which are protected by the exact gauge symmetry, while the 
fermion mass term (\ref{fermimass}) gives masses to all fermion fields
including gaugino. This asymmetry causes crucial effects on the 
quantum corrections as will be explained in the Sec.\ref{2p}.

\section{Calculational methods }\label{Sec:method2}

\subsection{Parameterization of bosonic fields and gauge fixing}\label{fixing}
In Ref.~\cite{Kaplan2}, bosonic fields $s,v$ are defined
by the real and imaginary parts of fluctuations of
$x,y$ from the (VEV) as in Eq.~(\ref{expansion}).
As pointed out in Ref.~\cite{Unsal},
we could instead take the following parameterization to define 
bosonic fields $s,v$:
\begin{align}
&x_\mathbf{n}=\frac{1}{\sqrt{2}}(\frac{1+\langle s \rangle}{a}+s_{x\mathbf{n}})
e^{iav_{x\mathbf{n}}}
\quad
y_\mathbf{n}=\frac{1}{\sqrt{2}}(\frac{1+\langle s \rangle}{a}+s_{y\mathbf{n}})
e^{iav_{y\mathbf{n}}}\label{y1}\\
&\bar{x}_\mathbf{n}=\frac{1}{\sqrt{2}}e^{-iav_{x\mathbf{n}}}
(\frac{1+\langle s \rangle}{a}+s_{x\mathbf{n}})
\quad
\bar{y}_\mathbf{n}=\frac{1}{\sqrt{2}}e^{-iav_{y\mathbf{n}}}
(\frac{1+\langle s \rangle}{a}+s_{y\mathbf{n}}),
\label{bar1}
\end{align}
where $\langle s \rangle$ represents the shift of the VEV
by quantum corrections. 
This parameterization is convenient since one can separate the 
gauge transformation property for $s$ and $v$; $s$ 
transform as adjoint site fields
under gauge transformation while $v$ transforms as 
bifundamental link variable.
In the following analysis we adopt the parameterization in 
Eqs.(\ref{y1}),~(\ref{bar1}).

We introduce the gauge fixing term:
\begin{equation}
S_{gf}=\frac{1}{2g^2}\{\frac{1}{\sqrt{2}}\alpha^g\}^2 
\sum_\mathbf{n} 
Tr[\{\nabla _{x}^-(x_\mathbf{n}-\bar{x}_\mathbf{n}) 
+\nabla _{y}^-(y_\mathbf{n}-\bar{y}_\mathbf{n})\}^2],
\label{gf}
\end{equation}
where
$\nabla^{\pm}_{x,y}$ 
are difference operators in the forward or backward
directions $\nabla^{\pm}_{x}f_{\mathbf{n}}
=\pm\frac{1}{a}\{f_{\mathbf{n}\pm \mathbf{i}}-f_{\mathbf{n}}\} $
,$\nabla^{\pm}_{y}f_{\mathbf{n}}
=\pm\frac{1}{a}\{f_{\mathbf{n}\pm \mathbf{j}}-f_{\mathbf{n}}\} $
 and 
$\alpha^g$ is an arbitrary parameter. 
We take $\alpha^g$ as 
the Feynman gauge $\alpha^g=(1+\langle s \rangle)$  which 
make the propagator 
of the gauge fields diagonal. 
From the gauge fixing condition (\ref{gf}) and the notation 
(\ref{y1},\ref{bar1}), ghost term is expressed as follows.

\begin{align}
S_{gh}=
&\frac{1}{2\bar{g}^2}(1+<s>) \sum_\mathbf{n} 
Tr[\bar{c}_\mathbf{n}(\nabla _{x}^-\{i[c_\mathbf{n},
(\frac{1+\langle s \rangle}{a}+s_{x\mathbf{n}})
U_{x\mathbf{n}}]-i(\frac{1+\langle s \rangle}{a}
+s_{x\mathbf{n}})U_{x\mathbf{n}}\nabla_{x}^+c_\mathbf{n}\nonumber\\
&-i[c_\mathbf{n},U^{\dagger}_{x\mathbf{n}}
(\frac{1+\langle s \rangle}{a}+s_{x\mathbf{n}})]
-i\nabla_{x}^+c_\mathbf{n}U^{\dagger}_{x\mathbf{n}}(\frac{1+\langle s 
\rangle}{a}+s_{x\mathbf{n}})\}\nonumber\\
&\nabla _{y}^-\{i[c_\mathbf{n},
(\frac{1+\langle s \rangle}{a}+s_{y\mathbf{n}})U_{y\mathbf{n}}]
-i(\frac{1+\langle s \rangle}{a}+s_{y\mathbf{n}})
U_{y\mathbf{n}}\nabla_{y}^+c_\mathbf{n}\nonumber\\
&-i[c_\mathbf{n},U^{\dagger}_{y\mathbf{n}}
(\frac{1+\langle s \rangle}{a}+s_{y\mathbf{n}})]
-i\nabla_{y}^+c_\mathbf{n}U^{\dagger}_{y\mathbf{n}}(\frac{1
+\langle s \rangle}{a}+s_{y\mathbf{n}})\}), 
\label{ghost}
\end{align} 
where $c_{\mathbf{n}},\bar{c}_{\mathbf{n}}$ are ghost and anti-ghost fields 
respectively, and 
$U_{\nu\mathbf{n}}=e^{iav_{\nu \mathbf{n}}}\, (\nu= x,y)$.

\subsection{Treatment of zero momentum modes}
\label{Sec:perturbation}

In the present theory the coupling $\bar{g}$ is the product of two-dimensional 
gauge coupling $g_2$ and lattice spacing $a$ as $\bar{g}=g_2a$. 
For a fixed gauge coupling $g_2$,  the dimensionless coupling $\bar{g}$ 
becomes small near the continuum limit, therefore the perturbation 
theory is expected to become a good approximation.
However, since there is no quadratic term of massless zero momentum modes, 
perturbative calculations
based on the gaussian integral becomes ill-defined, 
thus a special care must be taken for the zero momentum modes. 
In our approach, we carry out non-perturbative calculation for the zero 
momentum modes
while non-zero momentum modes are treated perturbatively.
\\

The calculational procedures are the following:
\begin{enumerate}
\item We perform 
the fourier transformation of the fields as given in Appendix.~A,
including the rescaling of the fields by certain powers in 
$\bar{g}$ and $N$.
$\bar{g},\bar{\mu},\frac{1}{N}$ are used as 
the parameter for perturbative expansion,
where 
$\bar{\mu}=a\mu=\frac{a}{L}$.

\item We carry out exact fermionic integral for both zero momentum modes and 
non-zero momentum modes. Then we also carry out 1-loop perturbation for the 
non-zero momentum bosonic fields. 

\item 
The effective action is the sum of the tree level action for the 
zero momentum bosonic fields and 
logarithm of the determinant from the 1-loop integral for
other fields which also depends on the zero momentum boson fields.
Expanding the 1-loop contribution in terms of the zero momentum 
boson fields, the leading term is a constant and next leading and 
next-to-next leading terms are the 1-point and 2-point functions. 
By the discussion in Sec.~\ref{Sec:method}, only these three terms 
can be relevant and higher terms in the effective action are irrelevant. 
We show that the 1-point and 2-point functions at the effective 
potential are irrelevant by explicit calculation, which 
will be described in Sec.~\ref{p3}

\item 
Once the  1-point and 2-point functions in the effective potential 
are shown to be irrelevant, these terms can be neglected in the 
effective action. Then the final form of the path-integral over the zero 
momentum bosonic fields can be reduced into a simpler form.
A non-perturbative calculation of the path-integral will be described 
in Sec.~\ref{nonp}.
\end{enumerate}

\section{Results} \label{Sec:non0}

\subsection{Procedure 1: Fourier transformation}\label{p1}

Let us consider the following 1-point and 2-point functions of the 
scalar fields 
$s_{\mu}=s_{\mu}^{\alpha}T^{\alpha}$, 
where $\mu=x,y$ and $T^{\alpha}(\alpha=0,1,2,3)$ are the generator of 
$U(2)$ gauge 
group with $T^0= \frac{1}{2}\times \mathbf{1}$ and
pauli matrices $T^a=\frac{1}{2}\sigma^a (a=1,2,3)$ .
\begin{eqnarray}
I^{\alpha}_1 &\equiv& \sum_{\mathbf{n}} 
\langle s^{\alpha}_{\mu \mathbf{n}} \rangle 
=  
\frac{\int     \displaystyle{\prod_{\beta,\mathbf{n}}}
      (\displaystyle{\prod_{\nu}}
      d \phi_{\nu\mathbf{n}}^{\beta})  
d \psi_{\mathbf{n}}^{\beta} d \bar{\psi}_{\mathbf{n}}^{\beta} 
      det'(D_{gh})
      \sum_{\mathbf{n}} s^{\alpha}_{\mu \mathbf{n}} 
        e^{-S}}
{\int    \displaystyle{\prod_{\beta,\mathbf{n}}}
      (\displaystyle{\prod_{\nu}}
      d \phi_{\nu\mathbf{n}}^{\beta}) 
d \psi_{\mathbf{n}}^{\beta}d \bar{\psi}_{\mathbf{n}}^{\beta} 
      det'(D_{gh})
        e^{-S}}
\nonumber\\
I^{\alpha,\beta}_2 &\equiv& 
\sum_{\mathbf{n}} \langle s^{\alpha}_{\mu \mathbf{n}} 
s^{\beta}_{\mu \mathbf{m}} \rangle 
=  
\frac{\int     \displaystyle{\prod_{\gamma,\mathbf{n}}}
      (\displaystyle{\prod_{\nu}}
      d \phi_{\nu\mathbf{n}}^{\gamma})  
d \psi_{\mathbf{n}}^{\gamma} d \bar{\psi}_{\mathbf{n}}^{\gamma} 
      det'(D_{gh})
      \sum_{\mathbf{n}} s^{\alpha}_{\mu \mathbf{n}} s^{\beta}_{\mu \mathbf{m}} 
        e^{-S}}
{\int    \displaystyle{\prod_{\gamma,\mathbf{n}}}(\displaystyle{\prod_{\nu}}
      d \phi_{\nu\mathbf{n}}^{\gamma})
      d \psi_{\mathbf{n}}^{\gamma} d \bar{\psi}_{\mathbf{n}}^{\gamma} 
      det'(D_{gh})
        e^{-S}}
\nonumber,
\label{patition}
\end{eqnarray}
where subscript $\alpha,\beta,\gamma$ stand for the 
$U(2)$ gauge generator and "$\phi_{\nu}^{\beta}$" 
in the integration measure 
are defined as 
\begin{align}
\phi_{\mu}=\phi_{\mu}^{\beta}T^{\beta}, (\mu=0,1,2,3)\\
\phi_{0}^{\beta}=s_x^{\beta}, 
\, \phi_{1}^{\beta}=-v_y^{\beta}, 
\, \phi_{2}^{\beta}=s_y^{\beta}, 
\, \phi_{3}^{\beta}=-v_x^{\beta}.
\end{align}
"$\psi^{\gamma}$" denotes the fermionic fields $\lambda^{\gamma}$ and 
$\xi^{\gamma}$ , 
and "$\bar{\psi}^{\gamma}$" denotes the fermionic fields 
$\alpha^{\gamma}$ and $\beta^{\gamma}$.
In the following we omit the subscript '$\mu$' of '$\phi_\mu$' 
when it is possible. 
$det'(D_{gh})$ is the Fadeev-Popov ghost determinant, 
where 
the contributions from the 
zero modes which correspond to the residual gauge symmetry are removed. 
The total action $S$ is $S=S_2+S_{meas}+S_{gf}$, where 
$S_{meas}$ is the measure term from the definition of notation 
(\ref{y1}),(\ref{bar1}), we describe the detailed discussion at that 
measure term on Appendix.~\ref{Sec:measure}. 
Here we represent the fields by momentum representation 
as described on (\ref{cano1})-(\ref{cano4a}). 
The above 1- and 2- point functions 
in momentum representation 
are
\begin{eqnarray}
I_n^{\alpha_1 \cdots \alpha_n}  = 
g_2^{\frac{n}{2}}L^{\frac{3n}{2}}
\frac{\int \displaystyle{\prod_{\mathbf{k}}} d\tilde{\phi}(\mathbf{k}) 
d\tilde{\psi}(\mathbf{k}) d\bar{\tilde{\psi}}(\mathbf{k} )
det'(D_{gh})
\displaystyle{\prod_{i=1}^n}
(\tilde{s}^{\alpha_i}_{\mu}(0))  e^{-S}}
{\int \displaystyle{\prod_{\mathbf{k}}} 
d\tilde{\phi}(\mathbf{k}) 
d\tilde{\psi}(\mathbf{k}) 
d\bar{\tilde{\psi}}(\mathbf{k}) 
det'(D_{gh})
        e^{-S}},
\label{momentumrep}
\end{eqnarray}
with $n=1,2$. 

The action is expressed in terms of the fourier modes as 
\begin{eqnarray}
S = S_b + S_f,
\end{eqnarray}
where the bosonic part $S_b$ is 
\begin{eqnarray}
S_b & = & \sum_{\mathbf{k}\neq 0}  \tilde{\phi}_{\mu}(\mathbf{k})
       D_{\phi}(\mathbf{k})^{\mu,\nu}
       \tilde{\phi}_{\nu}(-\mathbf{k}) 
+ S_{zero} + S_{meas} + O(\frac{\bar{g}}{N} \tilde{\phi} (\mathbf{k})^3),
\end{eqnarray}
where we have written the kinetic term symbolically as 
$D_{\phi}(\mathbf{k})$. 
We note that this kinetic term 
depends on the zero momentum modes of the bosonic fields $\tilde{\phi}(0)$.
$S_{zero}$ is the zero momentum mode part of the bosonic action given as
\begin{eqnarray}
S_{\rm zero}
& =  &
\frac{1}{2}\sum_{\mu >\nu}Tr [\tilde{\phi}_{\mu}(0), \tilde{\phi}_{\nu}(0)]^2
  + \frac{\bar{\mu}}{\bar{g}}
Tr[\Bigl(\tilde{s}_x(0)
+\sqrt{\frac{\bar{g}}{N}}\frac{\tilde{s}_x^2(0)
}{2}\Bigr)^2 
  + \Bigl(\tilde{s}_y(0)+\sqrt{\frac{\bar{g}}{N}}\frac{\tilde{s}_y^2(0)
}{2}\Bigr)^2 ],
\nonumber\\
\end{eqnarray}
The fermion action $S_f$ in the momentum representation is
\begin{eqnarray}
S_f 
&=& \left(
\begin{array}{cc|cc|c|c}
\bar{g}^{-1}N
 (\frac{\bar{g}}{N})^{1/2} \tilde{\alpha}_{\mathbf{k}}^\mu,& (\frac{\bar{g}}{N})^{1/2} \tilde{\beta}_{\mathbf{k}}^\mu,&
\tilde{\alpha}_{\mathbf{0}}^a, &
\tilde{\beta}_{\mathbf{0}}^a,&
\tilde{\beta}_{\mathbf{0}}^0, &
\tilde{\alpha}_{\mathbf{0}}^0 
\end{array}
\right)
\nonumber\\
 &\times&\left( \begin{array}{c|c|c|c}
  \bar{A}_{(2,2)\mathbf{k},\mathbf{p}}^{\mu\nu}
& \left( \frac{\bar{g}}{N} \right) B_{(2,2)\mathbf{k},\mathbf{0}}^{\mu b}
& \left( \frac{\bar{g}}{N} \right) C_{\xi (2,1)\mathbf{k},\mathbf{0}}^{\mu 0}
& \left( \frac{\bar{g}}{N} \right) \bar{\mu}_F 
  C_{\lambda (2,1)\mathbf{k},\mathbf{0}}^{\mu 0}
\\ \hline
  \left( \frac{\bar{g}}{N} \right) D_{(2,2)\mathbf{0},\mathbf{p}}^{a\nu}
& \bar{\mu}_F E'_{2,2}+
   \left( \frac{\bar{g}}{N} \right)^{\frac{1}{2}} 
   E_{(2,2)\mathbf{0},\mathbf{0}}^{ab}
& \left( \frac{\bar{g}}{N} \right)^{\frac{1}{2}}
    \bar{\mu}_F F_{\xi (2,1)\mathbf{0},\mathbf{0}}^{a0}
& \left( \frac{\bar{g}}{N} \right)^{\frac{1}{2}}\bar{\mu}_F 
   F_{\lambda (2,1)\mathbf{0},\mathbf{0}}^{a0} \\ \hline 
  \left( \frac{\bar{g}}{N} \right) G_{(1,2)\mathbf{0},\mathbf{p}}^{0\nu}
& \left( \frac{\bar{g}}{N} \right)^{\frac{1}{2}}
   \bar{\mu}_F H_{\lambda (2,1)\mathbf{0},\mathbf{0}}^{0b} 
& \bar{\mu}_F 
& -\bar{\mu}_F \\ \hline 
\left( \frac{\bar{g}}{N} \right)J_{(1,2)\mathbf{0},\mathbf{p}}^{0\nu}
& \left( \frac{\bar{g}}{N} \right)^{\frac{1}{2}}
   \bar{\mu}_FK_{(1,2)\mathbf{0},\mathbf{0}}^{0b} 
& \bar{\mu}_F 
& \bar{\mu}_F 
\end{array}
\right)
\left(
\begin{array}{c}
(\frac{\bar{g}}{N})^{1/2}\tilde{\lambda}_{\mathbf{p}}^\nu \\
(\frac{\bar{g}}{N})^{1/2}\tilde{\xi}_{\mathbf{p}}^\nu \\
\hline
 \tilde{\lambda}_{\mathbf{0}}^a \\
 \tilde{\xi}_{\mathbf{0}}^a \\
\hline
 \tilde{\xi}_{\mathbf{0}}^0 \\
\hline
 \tilde{\lambda}_{\mathbf{0}}^0
\end{array}
\right)
\nonumber\\
\label{fermatSec:3}
\end{eqnarray}
with $\bar{\mu}_F=a\mu_F$ 
, and the submatrices
$\bar{A}_{(2,2)\mathbf{k},\mathbf{p}}^{\mu\nu},B_{(2,2)\mathbf{k},\mathbf{0}}^{\mu b},\cdots
,K_{(1,2)\mathbf{0},\mathbf{0}}^{0b} $ are given 
in Appendix.~\ref{Sec:fermionmatrix}.

The ghost action $S_{gh}$ is 
\begin{eqnarray}
S_{gh} 
&=& \bar{g}^{-1}N
( (\frac{\bar{g}}{N})^{1/2} \bar{\tilde{c}}_{\mathbf{k}}^\mu, 
  \tilde{\bar{c}}_{\mathbf{0}}^a, 
  \tilde{\bar{c}}_{\mathbf{0}}^0 )
\left( \begin{array}{c|c|c}
  \bar{\Upsilon}_{\mathbf{k},\mathbf{p}}^{\mu\nu}
& \left( \frac{\bar{g}}{N} \right) \Theta_{\mathbf{k},\mathbf{0}}^{\mu b}
& 0 
\\ \hline
  0 & 0 & 0 \\ \hline 
  0 & 0 & 0 
\end{array}
\right)
\left(
\begin{array}{c}
(\frac{\bar{g}}{N})^{1/2}\tilde{c}_{\mathbf{p}}^\nu \\
 \tilde{c}_{\mathbf{0}}^a \\
 \tilde{c}_{\mathbf{0}}^0 
\end{array}
\right)
\label{ghmatSec:3}
\end{eqnarray}
$\bar{\Upsilon}_{\mathbf{k},\mathbf{p}}^{\mu\nu}$ and 
$\Theta_{\mathbf{k},\mathbf{0}}^{\mu b}$ are also given 
in Appendix.~\ref{Sec:fermionmatrix}.
\subsection{Procedure 2: 
Perturbative calculation of the non-zero momentum bosonic fields}\label{p2}
We now make 1-loop perturbation for the non-zero momentum bosonic fields.
It is easy to see that 
the 1-loop contribution is nothing but a Gaussian integral 
for the kinetic term and the contribution from the interaction terms 
gives higher order corrections, which can be neglected. This leaves only 
the determinant factor $det[ D_{\phi}(\mathbf{k}) ]^{-1/2}$ 
for the path-integral. This also simplifies the fermion path-integral. 
Since at 1-loop order in perturbation theory all the contributions 
of non-zero momentum bosonic fields can be dropped except for 
$det[ D_{\phi}(\mathbf{k}) ]^{-1/2}$ , 
the off-diagonal block parts in the fermion matrix 
$B_{(2,2)\mathbf{k},\mathbf{0}}^{\mu b}$, 
$C_{\xi (2,1)\mathbf{k},\mathbf{0}}^{\mu 0}$, 
$C_{\lambda (2,1)\mathbf{k},\mathbf{0}}^{\mu 0}$, 
$D_{(2,2)\mathbf{0},\mathbf{p}}^{a\nu}$, 
$G_{(1,2)\mathbf{0},\mathbf{p}}^{0\nu}$
and $J_{(1,2)\mathbf{0},\mathbf{p}}^{0\nu}$ can also dropped. 
It can be also shown that the matrix 
$\bar{A}_{(2,2)\mathbf{k},\mathbf{p}}^{\mu\nu}$ becomes 
$\bar{A}_{(2,2)\mathbf{k},\mathbf{p}}^{\mu\nu} 
=\bar{A}_{(2,2)\mathbf{k}}^{\mu\nu} \delta_{\mathbf{k},-\mathbf{p}} $. 
after dropping the non-zero momentum 
bosonic fields. 
Then it is easy to see that only the 
following term contributes to effective action in the determinant of 
the fermion matrix $M_f$. 
\begin{eqnarray}
det(M_f) 
&\propto& det( \bar{A}_{(2,2)\mathbf{k}}^{\mu\nu}) 
\nonumber\\
  &\times & det \left( \begin{array}{c|c|c}
 \bar{\mu}_F E'_{2,2}+
   \left( \frac{\bar{g}}{N} \right)^{\frac{1}{2}} 
   E_{(2,2)\mathbf{0},\mathbf{0}}^{ab}
& \left( \frac{\bar{g}}{N} \right)^{\frac{1}{2}}
    \bar{\mu}_F F_{\xi (2,1)\mathbf{0},\mathbf{0}}^{a0}
& \left( \frac{\bar{g}}{N} \right)^{\frac{1}{2}}\bar{\mu}_F 
   F_{\lambda (2,1)\mathbf{0},\mathbf{0}}^{a0} \\ \hline 
\left( \frac{\bar{g}}{N} \right)^{\frac{1}{2}}
   \bar{\mu}_F H_{\lambda (2,1)\mathbf{0},\mathbf{0}}^{0b} 
& \bar{\mu}_F 
& -\bar{\mu}_F \\ \hline 
\left( \frac{\bar{g}}{N} \right)^{\frac{1}{2}}
   \bar{\mu}_FK_{(1,2)\mathbf{0},\mathbf{0}}^{0b} 
& \bar{\mu}_F 
& \bar{\mu}_F 
\end{array}
\right)
\nonumber\\
& = & [ \prod_{k\neq 0} det[D_{\psi}(\mathbf{k})]] det[D_{\psi_0}].
\label{det}
\end{eqnarray}
The ghost determinant is given as
\begin{eqnarray}
det'[D_{gh}]=\prod_{k\neq 0} det[D_{gh}(\mathbf{k})] \equiv
\prod_{k\neq 0} det( \bar{\Upsilon}_{\mathbf{k}}^{\mu\nu}) .
\end{eqnarray}
We then obtain
\begin{align}
I_n^{\alpha_1 \cdots \alpha_n} &= 
L^{\frac{3n}{2}}g_2^{\frac{n}{2}}
\nonumber\\
&\times \frac{\int d\tilde{\phi}(\mathbf{0}) 
 det[D_{\psi}(\mathbf{0}) ]
\displaystyle{\prod_{\mathbf{k}\neq 0}}
\left( det[D_{\phi}(\mathbf{k}) ]^{-\frac{1}{2}}
det[D_{gh}(\mathbf{k})]
det[D_{\psi}(\mathbf{k}) ] \right)
\displaystyle{\prod_{i=1}^n}
(\tilde{s}^{\alpha_i}_{\mu}(0)) e^{-S_{\rm zero}}e^{-S_{meas}}}
{\int d\tilde{\phi}(\mathbf{0}) 
  det[D_{\psi}(\mathbf{0})]
\displaystyle{\prod_{\mathbf{k}\neq 0}}
\left( det[D_{\phi}(\mathbf{k}) ]^{-\frac{1}{2}}
det[D_{gh}(\mathbf{k}) ] 
det[D_{\psi}(\mathbf{k})] \right)
e^{-S_{\rm zero}}
e^{-S_{meas}}}. 
\label{patitiongauss}
\end{align}

\subsection{Procedure 3: Numerical study of the 1-loop contribution 
from the non-zero momentum modes }\label{p3}

We note here that 
$S_{meas}-log (\Pi_{\mathbf{k}}det[D_{\phi}(\mathbf{k})]^{-\frac{1}{2}}
det[D_{gh}(\mathbf{k})]det[D_{\psi}(\mathbf{k})])$ 
is nothing but the contribution from the non-zero momentum modes
to the 1-loop effective action 
and the zero momentum mode of measure term. 
In general, it depends on the 
zero momentum mode of the boson field. 
We expand the 1-loop effective action in the 
bosonic zero momentum mode, and effective action becomes as following. 
\begin{eqnarray}
S_{eff}
&=&
S_{zero}+S_{meas}
-log (\Pi_{\mathbf{k}}det[D_{\phi}(\mathbf{k})]^{-\frac{1}{2}}
det[D_{gh}(\mathbf{k})]det[D_{\psi}(\mathbf{k})])
\nonumber\\
&=& S_{zero}+S_{1loop}^{(0)}
+ (S_{zero}^{(1)}+S_{1loop}^{(1)})^{\alpha_1}_{\mu} 
\tilde{\phi}(\mathbf{0})^{\alpha_1}_{\mu}
+  (S_{zero}^{(2)}+S_{1loop}^{(2)})^{\alpha_1\alpha_2}_{\mu\nu} 
\tilde{\phi}(\mathbf{0})^{\alpha_1}_{\mu}
\tilde{\phi}(\mathbf{0})^{\alpha_2}_{\nu}
\nonumber\\
 && +(S_{zero}^{(3)}+S_{1loop}^{(3)})^{\alpha_1\alpha_2\alpha_3}_{\mu\nu\rho}
 \tilde{\phi}(\mathbf{0})^{\alpha_1}_\mu 
\tilde{\phi}(\mathbf{0})^{\alpha_2}_\nu 
\tilde{\phi}(\mathbf{0})^{\alpha_3}_\rho 
\nonumber\\
&&
+(S_{zero}^{(4)}
+S_{1loop}^{(4)})^{\alpha_1\alpha_2\alpha_3\alpha_4}_{\mu\nu\rho\sigma} 
\tilde{\phi}(\mathbf{0})^{\alpha_1}_{\mu}
\tilde{\phi}(\mathbf{0})^{\alpha_2}_{\nu}
\tilde{\phi}(\mathbf{0})^{\alpha_3}_{\rho}
\tilde{\phi}(\mathbf{0})^{\alpha_4}_{\sigma}
+\cdots ,
\label{Seff}
\end{eqnarray}
where $S_{1loop}^{(0)},S_{1loop}^{(1)},S_{1loop}^{(2)}...$ 
are coefficients derived by non-zero momentum mode integral and measure term. 
Among the contributions of 1-loop effective potential, only leading term and 
1-point and 2-point terms ($S_{1loop}^{(0)}$,
$(S_{1loop}^{(1)\alpha_1})_{\mu} \tilde{\phi}(\mathbf{0})^{\alpha_1}_{\mu}$,
$(S_{1loop}^{(2)\alpha_1\alpha_2})_{\mu\nu} 
\tilde{\phi}(\mathbf{0})^{\alpha_1}_{\mu}
\tilde{\phi}(\mathbf{0})^{\alpha_2}_{\nu}$)
can be relevant or 
marginal as suggested 
by the power counting in the previous section.

We now investigate whether 1- and 2-point functions from 1-loop contributions 
($S_{1loop}^{(1)}$, $S_{1loop}^{(2)}$) become irrelevant or not 
by the explicit calculation.
	
\subsubsection{1-point function}\label{1pper}
Due to the Furry's theorem and the gauge symmetry, the only 
fields which can have non-vanishing 
1-point functions 
are the $U(1)$ part of the scalar fields $\tilde{s}^{0}_{x,y}(0)$.
These 1- point functions can be absorbed into the shift of the 
VEV.
We represent the VEV 
which is proportional to the inverse lattice spacing including the 1-loop 
effect as $\frac{1+ \langle s \rangle }{\sqrt{2} a}$, 
where $\langle s \rangle$ 
corresponds to the shift of the VEV.

Using the expression of the measure term and fourier transformation 
in Appendices.~\ref{Sec:fourier} and \ref{Sec:measure},
we obtain 
the effective action for $U(M)$ gauge theory. 
For our explicit numerical calculation, we take $M=2$.
\begin{align}
S_{eff}(\langle s \rangle)|_{\tilde{\phi}(0)=0} 
=&\frac{\bar{\mu}^2}{2\bar{g}^2}
[(1+\langle s \rangle)^2-1]^2M\nonumber\\
&+\sum_{\mathbf{k}\neq 0}
\frac{1}{N^2} log[(1+\langle s \rangle)^2(\widehat{\mathbf{k}}^2+3\bar{\mu}^2)
-\bar{\mu}^2]M^2\nonumber\\
&-\sum_{\mathbf{k}\neq 0}\frac{1}{N^2} 
log[(1+\langle s \rangle)^2
\{\widehat{\mathbf{k}}^2(1+\bar{\mu_F})+2\bar{\mu_F}^2\}]M^2. \label{effec}
\end{align}

\begin{figure}
\begin{minipage}{.45\linewidth}
\includegraphics[width=6cm,angle=-90]{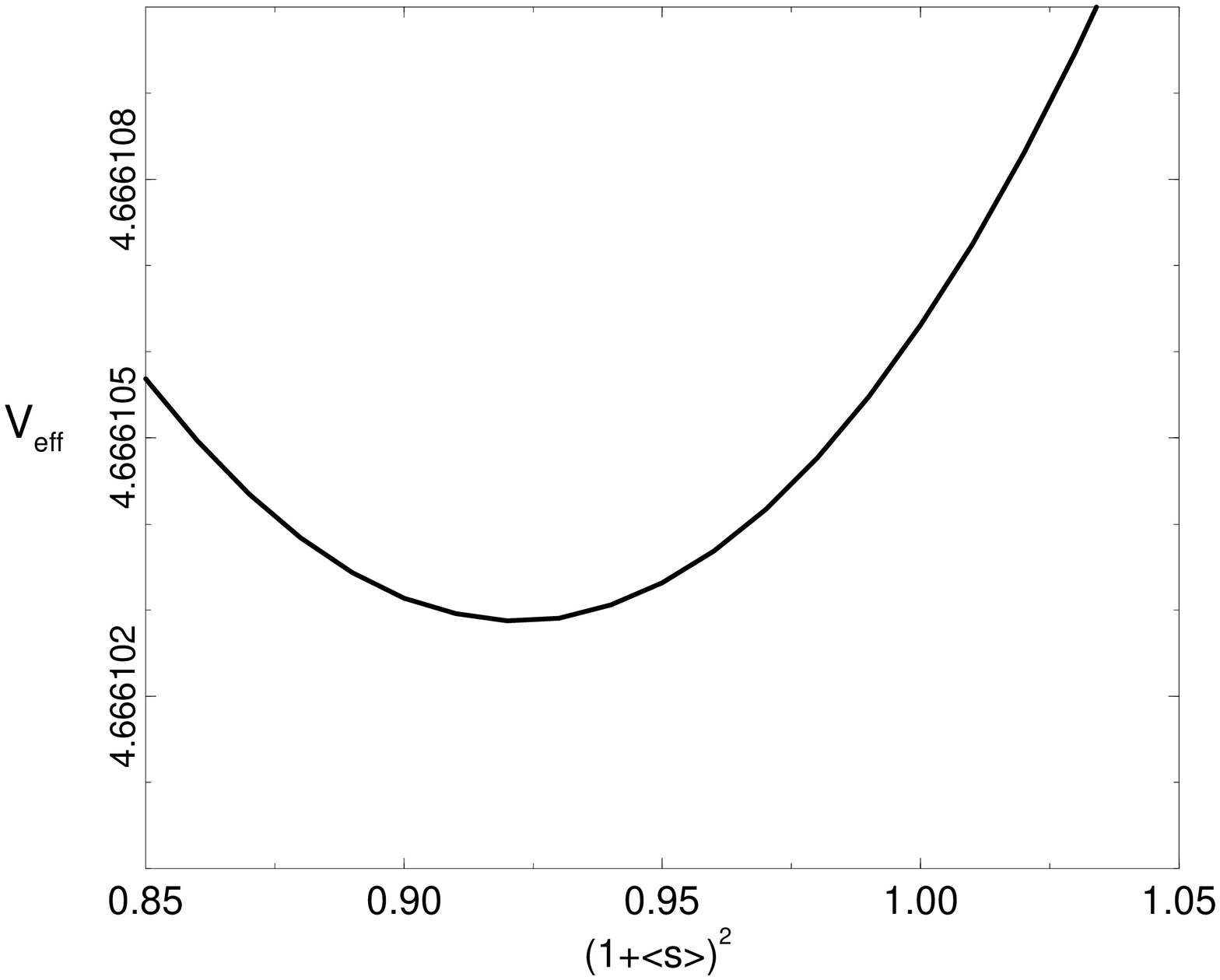}
\caption{The graph of $V_{eff}$ which depends on 
the 1-loop correction of lattice spacing $\frac{\langle s \rangle}{a}$.
Horizontal axis  is $(1+\langle s \rangle)^2$, 
Vertical one is $V_{eff}$.
We take parameters as 
$N=200,\bar{g}^2=\frac{3}{80}$ }
\label{N200g40}
\end{minipage}
\begin{minipage}{.06\linewidth}
.
\end{minipage}
\begin{minipage}{.45\linewidth}
\includegraphics[width=7cm,angle=-90]{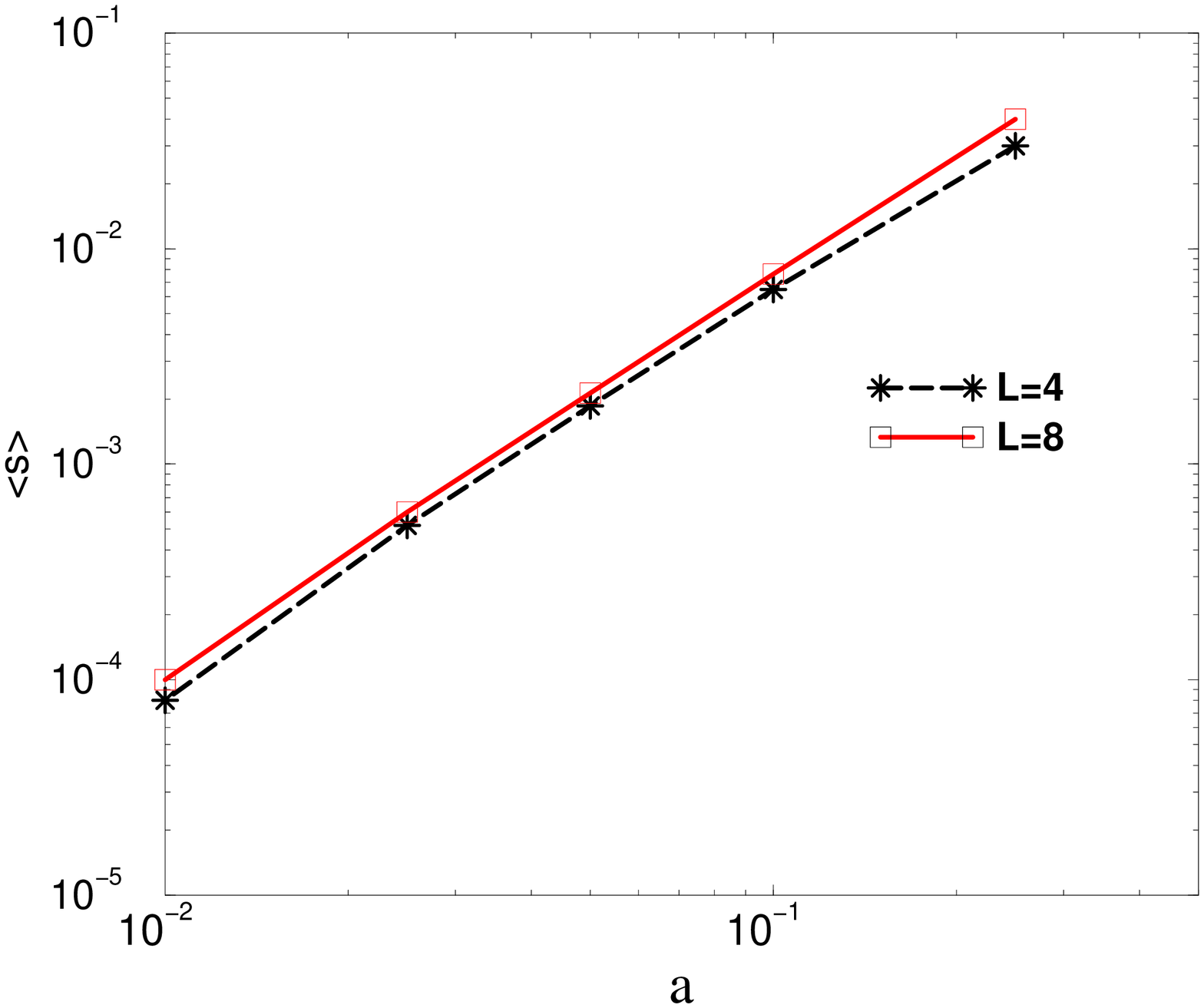}
\caption{
$a$ dependence of the global minima \\
$\langle s \rangle$ of $V_{eff}$.
Horizontal axis is lattice spacing $a$, Vertical one is 
$\langle s \rangle$. 
We take here $g_2=1$, 
the solid line is for volume $L=8$, while the dashed line is for $L=4$.
}
\label{hityou}
\end{minipage}
\end{figure}

For sufficiently small $\bar{g}=g_2 a$ 
we find that there is a minimum 
of the potential near $\langle s \rangle = 0$ as shown 
in Fig.~\ref{N200g40} where the 1-loop effective potential with the 
case $N=200$ and $\bar{g}^2=\frac{3}{80}$ is shown. 
The stability of the vacuum for more general parameter region will 
be studied in Sec.~\ref{decon}.
We also find that 
$\langle s \rangle$ vanishes quadratically in $a$ towards the continuum 
limit as shown in Fig.~\ref{hityou}.

\subsubsection{2-point function}\label{2p}

We next study whether 
the contribution from the non-zero momentum 
mode integral to the 2-point functions 
are relevant or not in the continuum limit. 
Among the 2-point terms 
$(S^{(2)}_{1-loop})^{\alpha_1\alpha_2}_{\mu\nu}
\phi^{\alpha_1}_{\mu}\phi^{\alpha_2}_{\nu}$ 
in the Eq.~(\ref{Seff}),
the terms of gauge fields are zero due to the gauge symmetry, and
only scalar 2-point terms for scalars $s_x(=\phi_0),s_y(=\phi_2)$ which 
are $(S^{(2)}_{1-loop})^{\alpha_1\alpha_2}_{00}
\phi^{\alpha_1}_{0}\phi^{\alpha_2}_{0}$ and 
$(S^{(2)}_{1-loop})^{\alpha_1\alpha_2}_{22}
\phi_{2}^{\alpha_1}\phi^{\alpha_2}_{2}$ are only non-zero. 
They are common due to the $Z_2$ symmetry between $x$ and $y$ directions.
Their analytical 
expressions are given in the Appendices.~\ref{ampli}.
In order to study the scaling properties of the ratio 
$S^{(2)}_{1-loop}/S^{(2)}_{zero}$,  it suffices to study 
$S^{(2)}_{1-loop}$ since the denominator has a fixed value
$\frac{\mu}{g_2}$ 
which does not depend on the lattice spacing.
The analytic form of $S^{(2)}_{1-loop}$ is
\begin{eqnarray}
(S^{(2)}_{1-loop})^{\alpha_1\alpha_2}_{\mu\nu}& = 
[&\delta_{\mu,0}\delta_{\nu,0}+\delta_{\mu,2}\delta_{\nu,2}]
[2\delta^{\alpha_1,0}\delta^{\alpha_2,0}S^{(2)}_{1-loop,~U(1)} 
+2M\delta^{\alpha_1,\alpha_2}S^{(2)}_{1-loop,~SU(2)} ]
\nonumber\\
S^{(2)}_{1-loop,~U(1)}& =
\frac{1}{2}[&\frac{1}{N^2}\sum_{\mathbf{k}}
\frac{2\widehat{\mathbf{k}}^2}
{(1+\langle s \rangle)^2(\widehat{\mathbf{k}}^2)^2}
+\frac{(F^{(1)}_{u1}+F^{(2)}_{u1})}
{(1+\langle s \rangle )^2
[\widehat{\mathbf{k}}^2(1+\bar{\mu}_F)+2\bar{\mu}_F^2]^2}
\nonumber\\
&&+\frac{-\frac{1}{2}\widehat{\mathbf{k}}^2+\frac{3}{4}\bar{\mu}^2}
{(1+\langle s \rangle)^2(\widehat{\mathbf{k}}^2+3\bar{\mu}^2)-\bar{\mu}^2}
+\frac{(1+\langle s \rangle)^2(\widehat{k_x}^4+\frac{3}{2}\bar{\mu}^2\widehat{\mathbf{k}}^2
+\frac{9}{2}\bar{\mu}^4)}
{[(1+\langle s \rangle)^2(\widehat{\mathbf{k}}^2+3\bar{\mu}^2)-\bar{\mu}^2]^2}]
\label{u1}
\nonumber\\
S^{(2)}_{1-loop,~SU(2)}&=\frac{1}{2}[&\frac{1}{N^2}\sum_{\mathbf{k}}
\frac{-2\widehat{\mathbf{k}}^2}{(1+\langle s \rangle)^2(\widehat{\mathbf{k}}^2)^2}
+\frac{(F^{(1)}_{su}+F^{(2)}_{su}+F^{(3)}_{su}+F^{(4)}_{su})}
{(1+\langle s \rangle )^2[\widehat{\mathbf{k}}^2(1+\bar{\mu}_F)+2\bar{\mu}_F^2]^2}
\nonumber\\
&&+\frac{\frac{3}{4}\bar{\mu}^2}
{(1+\langle s \rangle)^2(\widehat{\mathbf{k}}^2+3\bar{\mu}^2)-\bar{\mu}^2}
+\frac{(1+\langle s \rangle)^2(\widehat{k_x}^4+\frac{3}{2}\bar{\mu}^2\widehat{\mathbf{k}}^2
+\frac{9}{2}\bar{\mu}^4)}
{[(1+\langle s \rangle)^2(\widehat{\mathbf{k}}^2+3\bar{\mu}^2)-\bar{\mu}^2]^2}]
\label{ym},
\end{eqnarray}
where 
$S^{(2)}_{1-loop,~U(1)}$ and  $ S^{(2)}_{1-loop,~SU(2)} $
are the mass correction of the $U(1)$ and $SU(2)$ scalar fields.
$F^{(1)}_{u1},F^{(2)}_{u1}$ and 
$F^{(1)}_{su},F^{(2)}_{su},F^{(3)}_{su},F^{(4)}_{su}$ 
which appear in the fermion loop contributions to
$S^{(2)}_{1-loop,~U(1)}$ and  $ S^{(2)}_{1-loop,~SU(2)} $
are given as
\begin{align}
&F^{(1)}_{u1}=-(2\widehat{k_x}^2(1+\bar{\mu}_F)+\bar{\mu}_F^2)(1+\bar{\mu}_F)
\quad 
F^{(2)}_{u1}=-2[(1+\bar{\mu}_F)^2\widehat{k_y}^2-\bar{\mu}_F^2](1+\bar{\mu}_F)
\cos (k_xa) \nonumber\\
&F^{(1)}_{su}=\widehat{k_x}^2 \cos (k_xa)[(1+\bar{\mu}_F)^2+1] \quad
F^{(2)}_{su}=-\bar{\mu}_F(1+\bar{\mu}_F)^2\widehat{k_x}^2
+\bar{\mu}_F\widehat{k_x}\widehat{3k_x}\nonumber\\
&F^{(3)}_{su}=\bar{\mu}_F^2(1+\bar{\mu}_F)^2+\bar{\mu}_F\cos (2k_xa) \quad 
F^{(4)}_{su}= [(1+\bar{\mu}_F)^2\widehat{k_y}^2+\bar{\mu}_F^2]
[(1+\bar{\mu}_F)^2+1]
\end{align}

\begin{figure}
\begin{minipage}{.45\linewidth}
\includegraphics[angle=-90,width=9cm,clip]{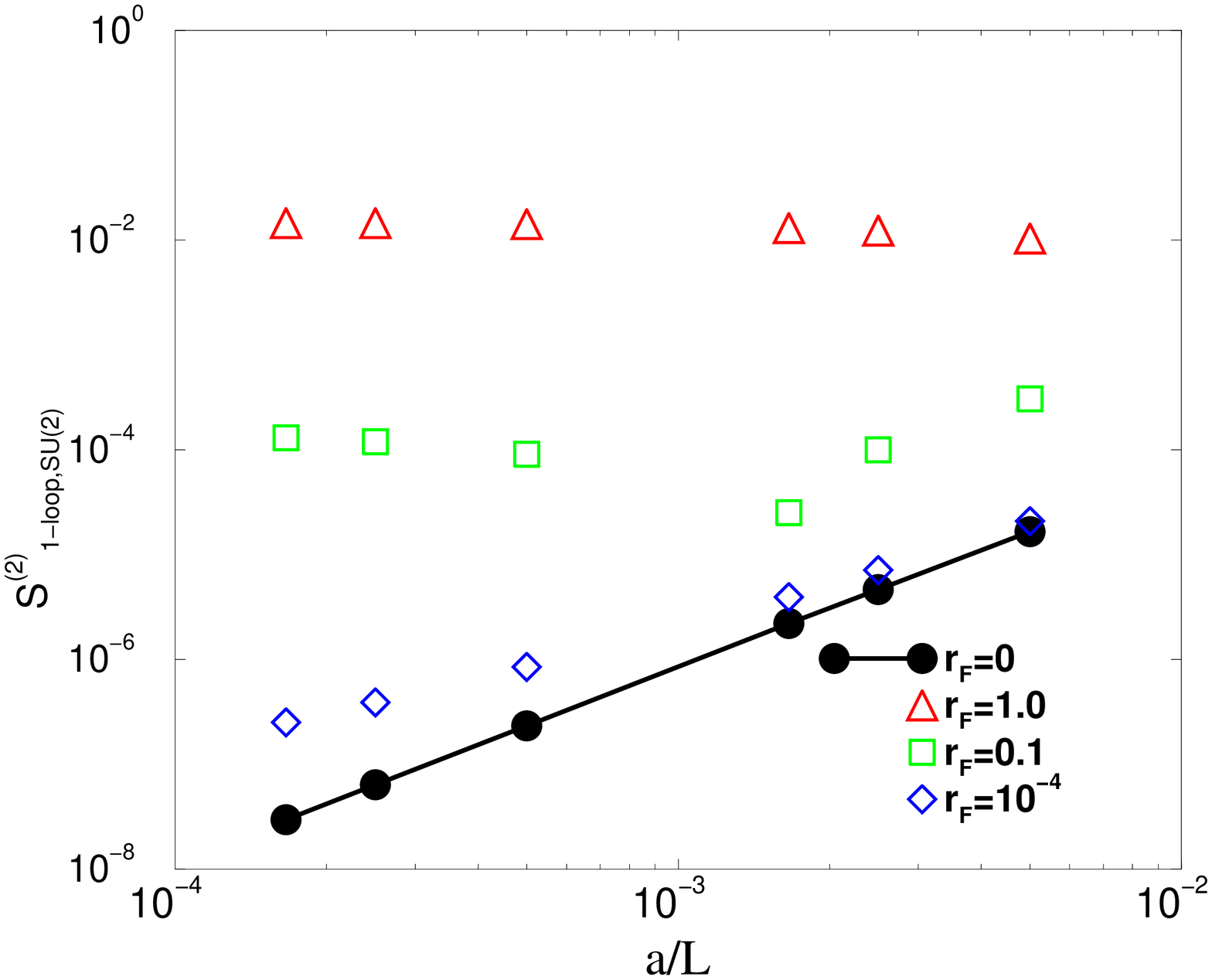}
\caption{($1/N$, $\bar{\mu}_F=r_F/N$) dependence of the nonabelian 
part of the 1-loop mass correction from the non-zero momentum mode. 
The horizontal axis is $\frac{1}{N}$ and the vertical axis is 
$S^{(2)}_{1-loop}$.}
\label{nonab}
\end{minipage}
\begin{minipage}{.05\linewidth}
.
\end{minipage}
\begin{minipage}{.45\linewidth}
\includegraphics[angle=-90,width=9cm,clip]{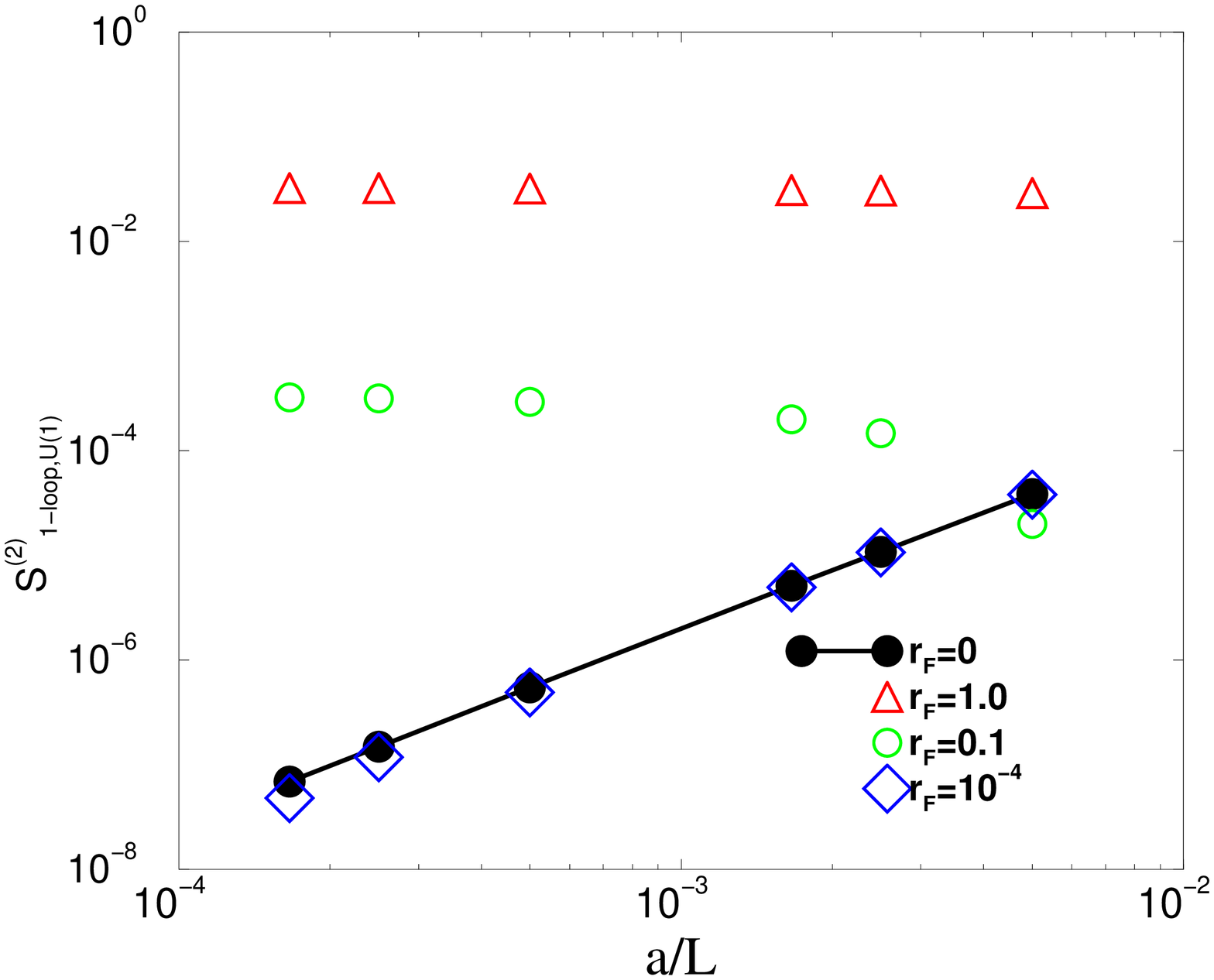}
\caption{($1/N$, $\bar{\mu}_F=r_F/N$) dependence of the abelian 
part of the 1-loop mass correction from the non-zero momentum mode. 
The horizontal axis is $\frac{1}{N}$ and the vertical axis is 
$S^{(2)}_{1-loop}$.}
\label{ab}
\end{minipage}
\end{figure}

\begin{figure}
\begin{minipage}{.45\linewidth}
\includegraphics[angle=-90,width=9cm,clip]{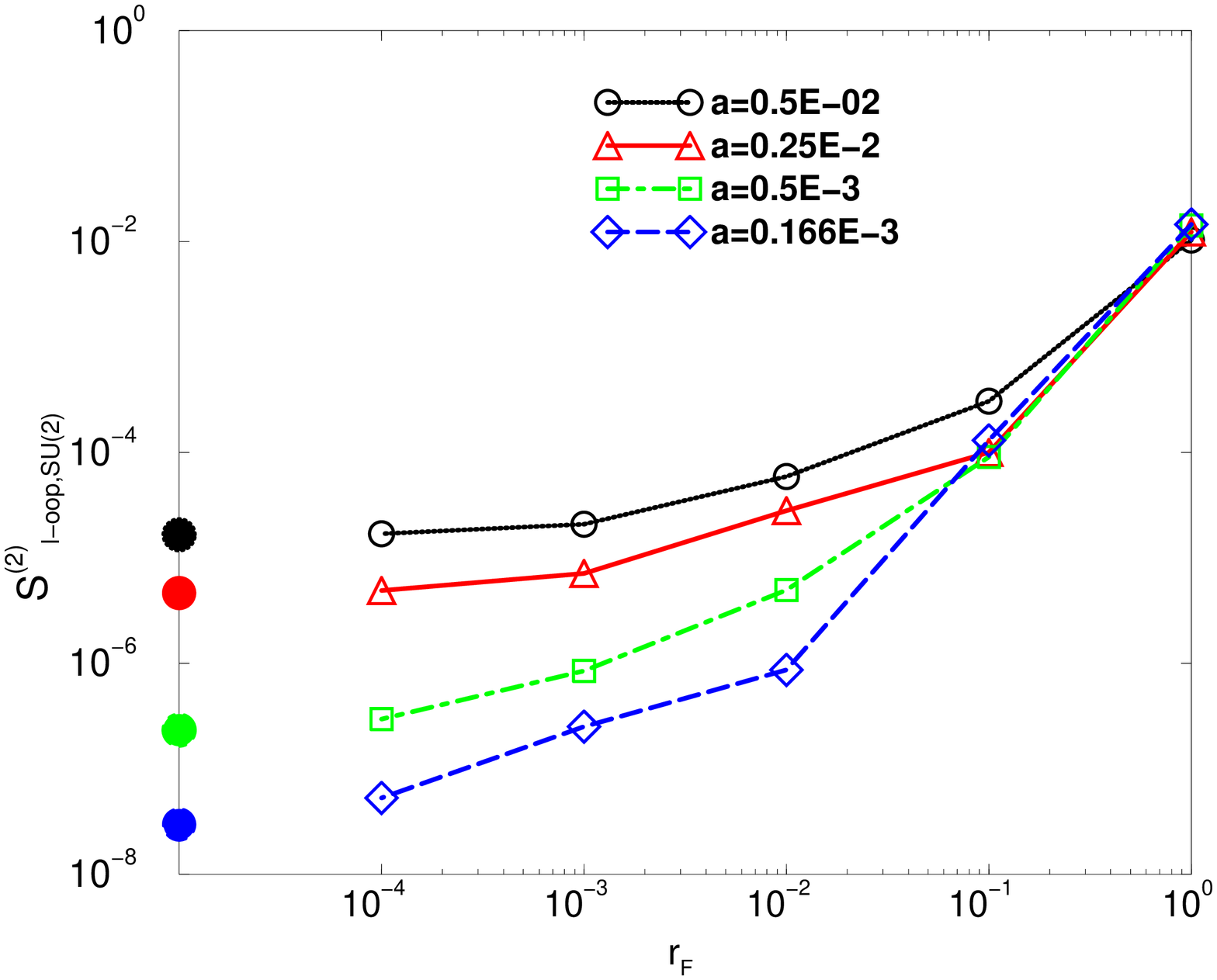}
\caption{($1/N$, $\bar{\mu}_F=r_F/N$) dependence of the nonabelian 
part of the 1-loop mass correction from the non-zero momentum mode. 
Horizon axis of this graph is $\frac{\mu_F}{\mu}=r_F$ 
and the vertical axis is 
$S^{(2)}_{1-loop}$.
Open symbols denote the data for non-zero $r_F$ for fixed 
lattice spacings.
In order to guide the eye, data for the same lattice spacing are 
connected by straight lines.
The filled circles are the values for $r_F=0$, for each lattice
spacing $a$.}
\label{alpvarasu2}
\end{minipage}
\begin{minipage}{.05\linewidth}
.
\end{minipage}
\begin{minipage}{.45\linewidth}
\includegraphics[angle=-90,width=9cm,clip]{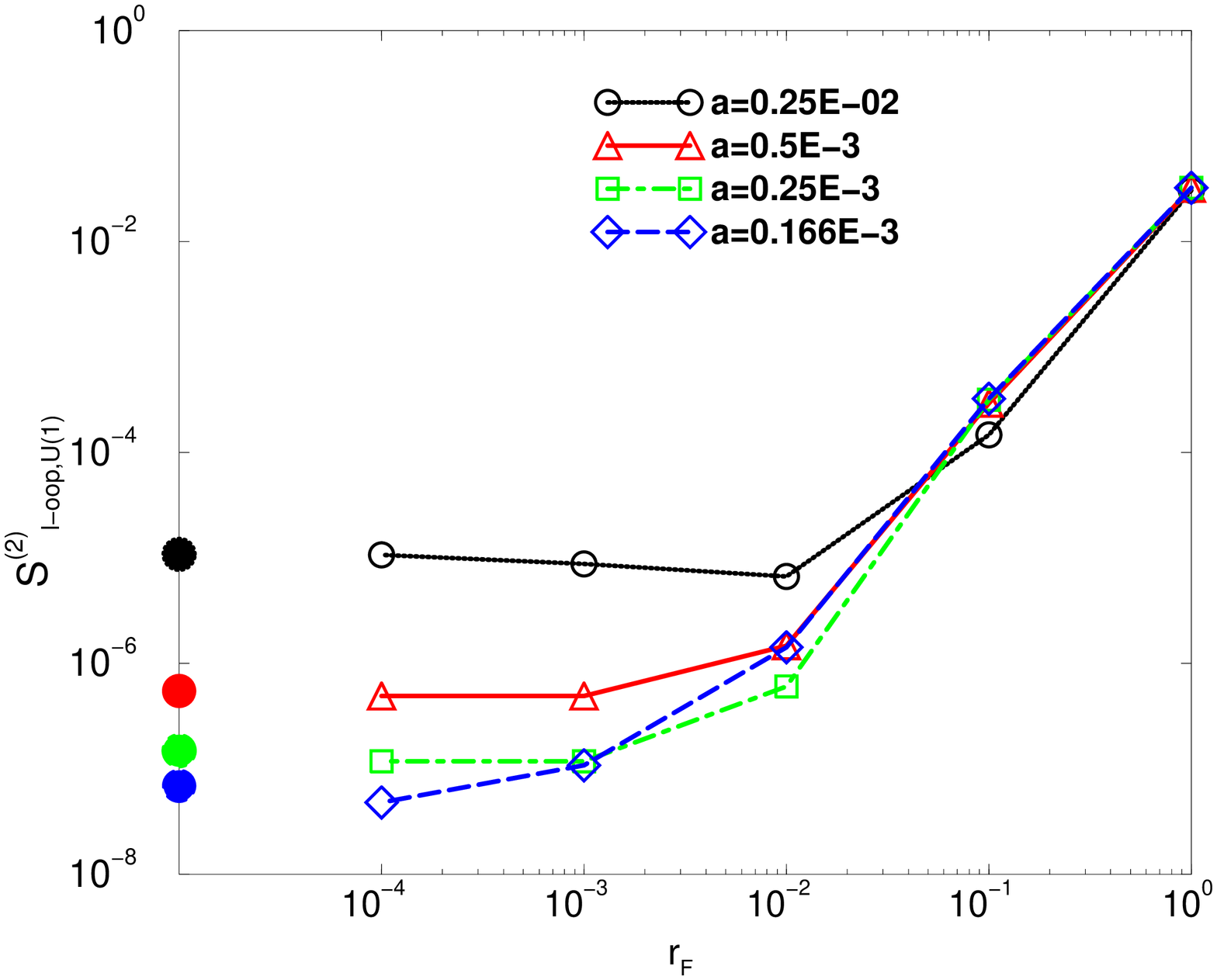}
\caption{($1/N$, $\bar{\mu}_F=r_F/N$) dependence of the abelian 
part of the 1-loop mass correction from the non-zero momentum mode. 
Horizon axis of this graph is $\frac{\mu_F}{\mu}=r_F$ 
and the vertical axis is 
$S^{(2)}_{1-loop}$.
Open symbols denote the data for non-zero $r_F$ for fixed 
lattice spacings.
In order to guide the eye, data for the same lattice spacing are 
connected by straight lines.
The filled circles are the values for $r_F=0$, for each lattice
spacing $a$.}
\label{alpvarau1}
\end{minipage}
\end{figure}

In order to see whether the 1-loop correction vanishes 
in the continuum limit,
we evaluate $S^{(2)}_{1-loop,~U(1)}$ $S^{(2)}_{1-loop,~SU(2)}$ 
in Eq.~(\ref{ym}) numerically.
The numerical 
results 
for several values of 
($1/N$, $\bar{\mu}_F\equiv r_F /N$) with $\bar{\mu}\equiv 1/N$ are given 
in Figs.~\ref{nonab} and \ref{ab}, with $r_F$ fixed.
Figs.~\ref{alpvarasu2}.~\ref{alpvarau1} shows the results for 
several values of $r_F$
with the lattice spacing $a$ fixed.

We find that the 1-loop correction 
for $r_F\neq 0$ does not vanish in the continuum limit, while 
that for $r_F=0$ vanishes. 
This scaling behavior can be understood as follows.
Let us divide the momentum integration region 
into two parts, i.e. high momentum parts:  $\mu \ll p \sim 1/a$, 
and low momentum parts:   $\mu \sim p \ll 1/a$. 
In high momentum region $\mu$ is a small perturbation 
and the leading contribution vanishes due to exact susy,
while the sub-leading contributions are suppressed by powers in $a$.
In low momentum region $\mu$ is not a small perturbation 
but the integral can be approximated by the continuum expression,
e.g. $\sin(p_{\mu}a) \to ap_{\mu} $, 
$\cos(p_{\mu}) \to 1 + O((ap)^2)$ etc.
Then the sum can be approximated by the integral with the infra-red 
cutoff $\bar{k}_0\equiv k_0a \sim 1/N$ and some 
intermediate ultra-violet cutoff $\bar{k}_1\equiv k_1a 
(\bar{k}_0\ll \bar{k}_1 \ll 1)$:
\begin{eqnarray}
S^{(2)}_{1-loop}
& \sim 2\delta^{\alpha_1,0}\delta^{\alpha_2,0} 
\displaystyle \int_{\bar{k}_0 \le |\bar{\mathbf{k}}| \le \bar{k}_1} 
\frac{d^2\bar{k}}{(2\pi)^2} \biggl[
& \frac{2\bar{\mathbf{k}}^2}{(\bar{\mathbf{k}}^2)^2}
+\frac{-2 \bar{\mathbf{k}}^2 +\bar{\mu}_F^2
}
{[\bar{\mathbf{k}}^2+2\bar{\mu_F}^2]^2}
\nonumber\\
& &
+\frac{-\frac{1}{2}\bar{\mathbf{k}}^2+\frac{3}{4}\bar{\mu}^2}
{\bar{\mathbf{k}}^2+2\bar{\mu}^2}
+\frac{\bar{k}_x^4+\frac{3}{2}\bar{\mu}^2 \bar{\mathbf{k}}^2
+\frac{9}{2}\bar{\mu}^4)}
{[\bar{\mathbf{k}}^2+2\bar{\mu}^2]^2}\biggr]
\label{u12}
\nonumber\\
& + 
2M\delta^{\alpha_1,\alpha_2} \displaystyle \int_{\bar{k}_0 \le 
|\bar{\mathbf{k}}| \le \bar{k}_1} 
\frac{d^2\bar{k}}{(2\pi)^2} \biggl[
& \frac{-2\bar{\mathbf{k}}^2}{(\bar{\mathbf{k}}^2)^2}
+\frac{2\bar{\mathbf{k}}^2 +\bar{\mu}_F + 2\bar{\mu}_F^2}
{[\bar{\mathbf{k}}^2+2\bar{\mu}_F^2]^2}
\nonumber\\
& & +\frac{\frac{3}{4}\bar{\mu}^2}{\bar{\mathbf{k}}^2+2\bar{\mu}^2}
+\frac{\bar{k}_x^4+\frac{3}{2}\bar{\mu}^2\bar{\mathbf{k}}^2
+\frac{9}{2}\bar{\mu}^4}{[\bar{\mathbf{k}}^2+2\bar{\mu}^2]^2} \biggr]
\label{ym2},
\end{eqnarray}
where we have set $\langle s \rangle = 0 $ for simplicity.
Then the first two terms in each integration 
give rise to contributions 
linear and logarithmic in the infra-red cutoff as
\begin{eqnarray}
S^{(2)}_{1-loop}
& \sim 2\delta^{a_1,0}\delta^{a_2,0} \biggl[
& \frac{1}{4\pi} ( 2 log(\frac{\bar{k}_1^2}{\bar{k}_0^2})
               - 2 log(\frac{\bar{k}_1^2+2\bar{\mu}_F^2}
                            {\bar{k}_0^2+2\bar{\mu}_F^2})
                +\frac{5\bar{\mu}_F^2}{\bar{k}_0^2+2\bar{\mu}_F^2} )\biggr]
\label{u13}
\nonumber\\
& + 
2M\delta^{a_1,a_2}\biggl[
& \frac{1}{4\pi} ( - 2 log(\frac{\bar{k}_1^2}{\bar{k}_0^2})
                + 2 log(\frac{\bar{k}_1^2+2\bar{\mu}_F^2}
                            {\bar{k}_0^2+2\bar{\mu}_F^2})
               +\frac{\bar{\mu}_F-2\bar{\mu}_F^2}{\bar{k}_0^2+2\bar{\mu}_F^2} )
\biggr]
\label{ym3}.
\end{eqnarray}
which give volume independent mass terms in the continuum limit.
One might naively wonder why setting $\mu = \mu_F$ and taking the
limits 
(1) $a \to 0$ 
then (2) $1/L \to 0$ does not work. This is because the contributions 
from infra-red parts are not completely canceled out
due to the asymmetry between the infra-red regulator 
of boson and fermion as mentioned in Sec.~\ref{Sec:method}, 
although the contributions from the UV part are canceled as expected.
\\
\\
In order to avoid the appearance of such counter terms, 
one should adopt the following procedure: 
\begin{enumerate}
\item Compute physical quantities for fixed ($1/N$, $\bar{\mu}_F=r_F/N$).
\item Take $\mu_F \to 0$ with fixed $1/N$ first , i.e $r_F \to 0$.
\item Then take the continuum limit, i.e.  $1/N \to 0$ .
\end{enumerate}
This two-step limit can avoid the counter terms as can be seen from 
Eq.~(\ref{u13}).

\subsection{Procedure 4: non-perturbative study of the zero momentum mode}
\label{nonp}
From the results of 
procedure 3
in the previous section, 
no term of 1-loop contributions 
from non-zero momentum modes 
to the effective action in Eq.~(\ref{Seff})
can survive 
in the continuum limit. 
Therefore in order to evaluate 
1- and 2-point functions in the continuum limit,
we only have to perform the following integral
\begin{align}
I_n^{\alpha_1 \cdots \alpha_n}  = 
L^{\frac{3n}{2}}g_2^{\frac{n}{2}}
\frac{\int d\tilde{\phi}(\mathbf{0}) 
 det[D_{\psi}(\mathbf{0}) ]
\displaystyle{\prod_{i=1}^n}
(\tilde{s}^{\alpha_i}_{\mu}(0)) e^{-S_{fin}}}
{\int d\tilde{\phi}(\mathbf{0}) 
 det[D_{\psi}(\mathbf{0}) ]
e^{-S_{fin}}}, \label{patition3}\\
S_{fin}
 =  \sum_{\mu >\nu}Tr [\tilde{\phi}_{\mu}(0), \tilde{\phi}_{\nu}(0)]^2
  + \frac{\mu}{g_2}
Tr[ (\tilde{s}_x(0)^2
+ \tilde{s}_y(0)^2) ].
\label{sfin}
\end{align}
We discuss whether the fine-tuning is needed or not 
by the investigation of the infinite volume behavior of this value.

To calculate 
Eq.~(\ref{patition3}), 
we should first 
express the fermion determinant 
$det[D_{\psi}(\mathbf{0})]$ from the zero momentum modes
as the function of 
bosonic fields analytically.
In the integral with only zero momentum bosonic modes, 
we can ignore the coordinate indices
$\mathbf{n}$, 
As is obvious from Eq.~(\ref{det}) 
the fermion determinant 
$det[D_{\psi}(\mathbf{0})]$ is given as
\begin{eqnarray}
det[D_{\psi}(\mathbf{0})]
  &= \bar{\mu}_F^2 det \left( \begin{array}{c|c|c}
  \bar{\mu}_F E'_{2,2}+
   \left( \frac{\bar{g}}{N} \right)^{\frac{1}{2}} 
   E_{(2,2)\mathbf{0},\mathbf{0}}^{ab}
& \left( \frac{\bar{g}}{N} \right)^{\frac{1}{2}}
    \bar{\mu}_F F_{\xi (2,1)\mathbf{0},\mathbf{0}}^{a0}
& \left( \frac{\bar{g}}{N} \right)^{\frac{1}{2}}\bar{\mu}_F 
   F_{\lambda (2,1)\mathbf{0},\mathbf{0}}^{a0} \\ \hline 
\left( \frac{\bar{g}}{N} \right)^{\frac{1}{2}}
   H_{\lambda (2,1)\mathbf{0},\mathbf{0}}^{0b} 
& 1
& -1 \\ \hline 
\left( \frac{\bar{g}}{N} \right)^{\frac{1}{2}}
    K_{(1,2)\mathbf{0},\mathbf{0}}^{0b} 
& 1
& 1
\end{array}
\right).
\nonumber\\
\end{eqnarray}
If one takes $\mu_F \to 0$ limit as the first part of the two step limit,
which was explained in Sec.~\ref{p3}
the determinant is simplified to a determinant of the $SU(2)$ group
\begin{eqnarray}
det[D_{\psi}(\mathbf{0})]
  \sim 2 \left(\frac{\bar{g}}{N} \right)^{\frac{1}{2}} 
     \bar{\mu}_F^2 
   det ( E_{(2,2)\mathbf{0},\mathbf{0}}^{ab} ),  
\end{eqnarray}
which is the fermion determinant of the following $SU(2)$ matrix model.
\begin{equation}
S=Tr(\frac{1}{2}\sum_{\mu> \nu}[\tilde{\phi}_{\mu},\tilde{\phi}_{\nu}]^2
+ \bar{\psi}\gamma_\mu[\tilde{\phi}_\mu, \psi]),
\label{matri}
\end{equation}
where
\begin{align}
\tilde{\phi}_{\mu}=\tilde{\phi}_{\mu}^aT^a, (\mu=0,1,2,3)\\
\tilde{\phi}_{0}^a=\tilde{s}_x^a, \, 
\tilde{\phi}_{1}^a=-\tilde{v}_y^a, \, \tilde{\phi}_{2}^a=\tilde{s}_y^a, \, 
\tilde{\phi}_{3}^a=-\tilde{v}_x^a.
\end{align}
with $T^a$s being $SU(2)$ generators. 
This action is invariant under $SO(4)$ Lorentz transformation
\begin{equation}
\tilde{\phi}_\mu \to \Lambda^{\nu }_{ \mu }\tilde{\phi}_{\nu }, 
\end{equation}
where 
$(\Lambda)^{\mu }_{\nu }$ is the $SO(4)$ matrix.  

The explicit form 
of the fermion 
determinant $det[D_{\psi}(\mathbf{0})]$ is 
\begin{align}
det[D_{\psi}(\mathbf{0})]=&\frac{1}{\bar{g}^{12}}
[(\tilde{\phi}^1\cdot \tilde{\phi}^1)(\tilde{\phi}^2\cdot \tilde{\phi}^2)
(\tilde{\phi}^3\cdot \tilde{\phi}^3)\nonumber\\
&-(\tilde{\phi}^1\cdot \tilde{\phi}^1)(\tilde{\phi}^2\cdot \tilde{\phi}^3)^2
-(\tilde{\phi}^2\cdot \tilde{\phi}^2)(\tilde{\phi}^1\cdot \tilde{\phi}^3)^2
-(\tilde{\phi}^3
\cdot \tilde{\phi}^3)(\tilde{\phi}^1\cdot \tilde{\phi}^2)^2\nonumber\\
&+2(\tilde{\phi}^1\cdot \tilde{\phi}^2)(\tilde{\phi}^2\cdot \tilde{\phi}^3)
(\tilde{\phi}^3\cdot \tilde{\phi}^1)]=
\frac{1}{\bar{g}^{12}}det(\tilde{\phi}^a\cdot \tilde{\phi}^b),
\label{ferd}
\\
&(\tilde{\phi}^a\cdot\tilde{\phi}^b\equiv \sum_{\mu= 0}^3
\tilde{\phi}^a_{\mu}
\tilde{\phi}^b_{\mu}),
\end{align}
which can be obtained 
as in Ref.~\cite{Suyama-Tsuchiya}. 

From (\ref{sfin}) and (\ref{ferd}), 
one can see that 
the fermion determinant and the bosonic action are even functions
in $\tilde{\phi}_{\mu}$. 
Since the 1-point function is odd in $\tilde{\phi}_{\mu}$, the integration of 
numerator of Eq.~(\ref{patition3}) for $n=1$ case trivially vanishes.

We now carry out the integral over the bosonic zero momentum mode 
in Eq.~(\ref{patition3}) for 2-point function non-perturbatively. 
We can decompose the action $S_{fin}$ as 
$S_{fin}=S_{SU(2)}+S_{U(1)}$, where 
$S_{SU(2)}$ and $S_{U(1)}$ are the actions for the $SU(2)$ 
and $U(1)$ part as
\begin{align}
&S_{SU(2)}=\sum_{\mu >\nu}Tr [\tilde{\phi}_{\mu}(0), \tilde{\phi}_{\nu}(0)]^2,
  + \frac{\mu}{g_2}
[ (\tilde{s}^a_x(0))^2
+ (\tilde{s}^a_y(0))^2 ]
,\\
&S_{U(1)}=\frac{\mu}{g_2}
[(\tilde{s}^0_x(0))^2
+ (\tilde{s}^0_y(0))^2].\label{sep}
\end{align}
Thus  the 2-point function in Eq.~(\ref{patition3}) can be 
factorized into the product of integrals over $U(1)$ fields and 
$SU(2)$ fields.
Since the fermion determinant is independent of the $U(1)$ part of the 
scalar fields, the $U(1)$ part of the 2-point function 
$I_2^{0,0}$
becomes a trivial gaussian integral and is identical to
the tree level value $g_2L^4$. 
Therefore only the $SU(2)$ part of the 2-point function $I_2^{a,b}$ 
becomes nontrivial as
\begin{align}
I^{a,b}_2 = g_2L^3
\frac{\int d\tilde{s}_x^a(\mathbf{0})d\tilde{v}_x^a(\mathbf{0})
d\tilde{s}_y^a(\mathbf{0})d\tilde{s}_y^a(\mathbf{0})
(\tilde{s}_{\mu}(0))^a(\tilde{s}_{\mu}(0))^b det[D_{\psi}(\mathbf{0}) ] 
e^{-S_{SU(2)}}}
{\int d\tilde{s}_x^a(\mathbf{0})d\tilde{v}_x^a(\mathbf{0})
d\tilde{s}_y^a(\mathbf{0})d\tilde{s}_y^a(\mathbf{0}) det[D_{\psi}(\mathbf{0}) ] e^{-S_{SU(2)}}} 
\equiv \delta^{a,b}\langle ss \rangle
\label{nonabelian}.
\end{align}
Since $\langle ss \rangle $ is the zero momentum mode of the propagator,
it can be written by the renormalized mass squared $m_R^2$
which is the sum of the tree level mass squared $\frac{\mu^2}{g_2^2}$ 
and the quantum correction $\Delta \mu^2$.
\begin{align}
\frac{1}{L^2}\langle ss \rangle =
\frac{1}{m_R^2}=
\frac{1}{\frac{\mu^2}{g_2^2}+\Delta\mu^2}\quad 
\label{masscore}
\end{align}
If there is no quantum correction, the 
2-point function becomes the tree level value $\langle ss \rangle_{tree}$ 
with $L$ dependence
\begin{align}
\frac{1}{L^2}\langle ss \rangle_{tree}
=\left( \frac{\mu^2}{g_2^2} \right)^{-1}=g_2L^2.
\end{align}

\subsubsection{Numerical calculation of the 2-point function } \label{num}
We perform the integral in Eq.~(\ref{nonabelian}) numerically.
Simulations are carried out in the  Metropolis algorithm
with $2.0 \times 10^{5}$ sweeps for the thermalization and 
$2.0 \times 10^{7}$ sweeps for the measurement. 
We estimate the error by the variance with binsize of $100$ sweeps.

\begin{figure}
\includegraphics[width=7cm,angle=-90]{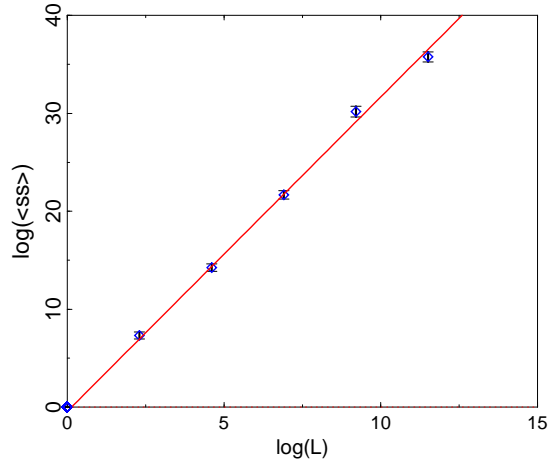}
\caption{The lattice size $L$ dependence of the 2-point function. 
The horizontal axis is $L$, where as the vertical axis is $\langle ss \rangle$. }
\label{0mode}
\end{figure}

Since the 2-point function depends only on the product $g_2 L$, 
we take $g_2=1$ without loosing generality. 
Fig.~\ref{0mode} shows the $L$ dependence of the 
2-point function 
$\langle ss \rangle$. 
As can be seen in Fig.~\ref{0mode}, we find that 
$\langle ss \rangle$ increases with $L$.
Fitting the data with the following function
\begin{equation}
\langle ss \rangle \sim  AL^{3+\alpha}, 
\label{numericalresult}
\end{equation}
we obtain $A=0.65(20)$ and $\alpha=0.210(46)$.
This gives the $L$ dependence of 
the renormalized mass 
\begin{equation}
m_R^2\equiv \frac{\mu^2}{g_2^2}+\Delta \mu^2 
\equiv L^2\langle ss \rangle^{-1} \sim  \frac{1}{AL^{1+\alpha}}, 
\label{mr}
\end{equation}
which vanishes in the large volume limit $L \to \infty$. 
Our result also implies that the contribution from the quantum 
corrections becomes dominant for large 
$L$.
Thus in the  continuum limit for finite volume, there is a 
non-trivial mass correction which is larger than the tree level contribution 
$\frac{\mu}{g_2}$. However, after taking the infinite volume limit 
the mass term vanishes so that there is no need for fine-tuning.

\section{Constraint from the stability of the lattice spacetime}
\label{decon}
In this section we study the stability of the lattice spacetime by 
the deconstruction against quantum effects.
In Sec.~\ref{1pper},
we found that with sufficiently small $\bar{g}$ and fixed $L$, 
there is a minimum of the 1-loop potential $V(\langle s \rangle)$ near 
the tree-level value and that the 1-loop shift of the expectation value 
vanishes towards the 
continuum limit so that the quantum correction becomes irrelevant. 
However, this may not always be the case for any choices of the parameters.
In general the tree level contribution of the potential is proportional 
to $\frac{\mu}{g_2}\sim \frac{1}{g_2 L}$, whereas the 1-loop correction 
depends on $\bar{\mu}$. If $g_2 L$ is too large 
there is a possibility that global minima may disappear and the 
lattice spacetime structure can be destroyed due to large quantum 
effects.
Therefore it is quite important to investigate 
in the parameter region of interest where the physical correlation 
length $g_2^{-1}$ is larger than the lattice spacing $a$ but smaller 
lattice size $L$,
\begin{equation}
a \ll (g_2)^{-1} \ll L,
 \label{physreg1}
\end{equation}

\begin{figure}
\begin{minipage}{.45\linewidth}
\includegraphics[width=6cm,angle=-90]{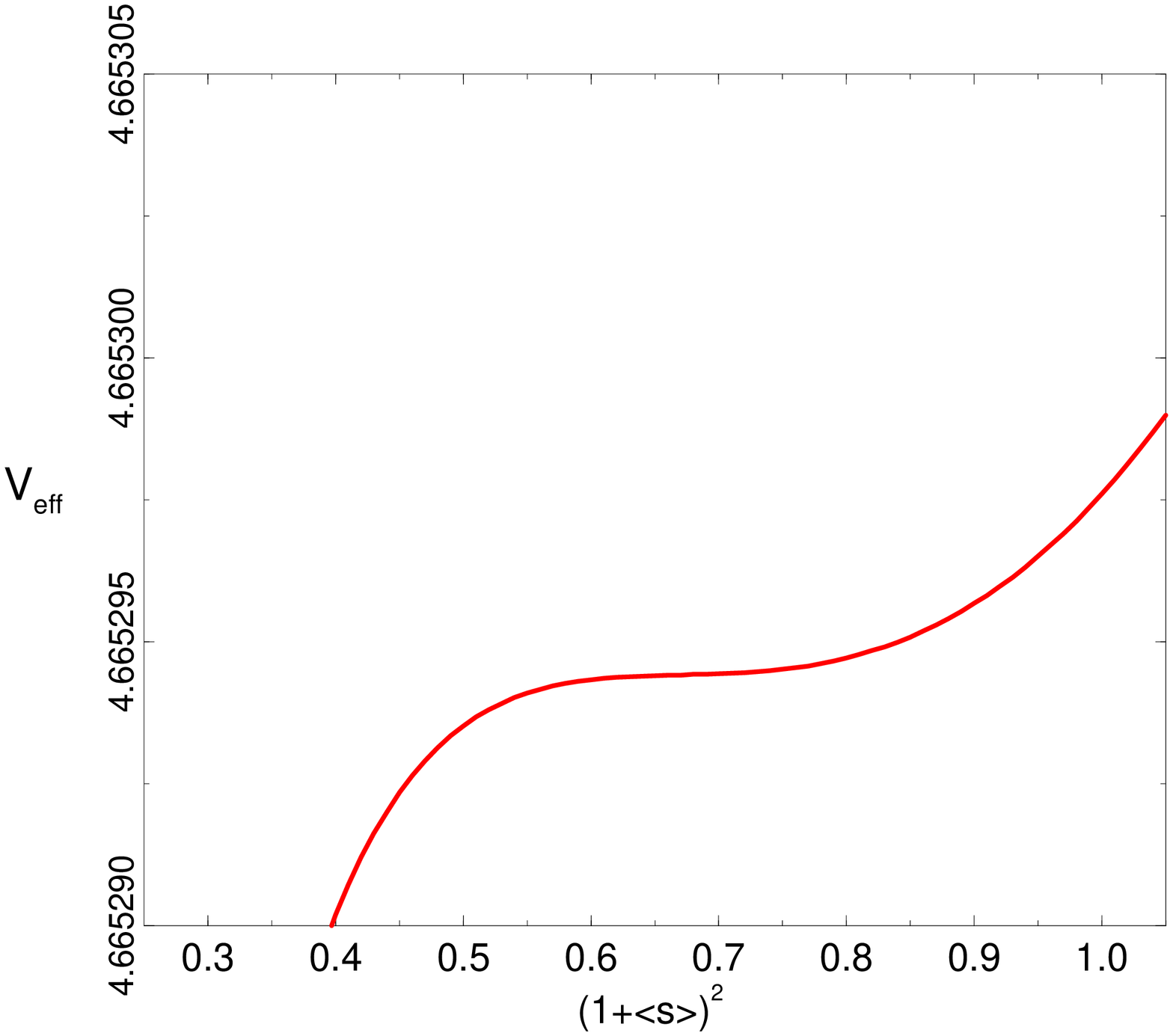}
\caption{The graph of 
$\langle s \rangle$dependence of $V_{eff}$ for $N=400,\bar{g}^2=0.075 \gtrsim 
0.2$.
\\ 
The horizontal axis  is $(1+\langle s \rangle)^2$, while the vertical
axis 
is $V_{eff}$.
}
\label{ng400}
\end{minipage}
\begin{minipage}{.05\linewidth}
.
\end{minipage}
\begin{minipage}{.45\linewidth}
\includegraphics[width=6cm,angle=-90]{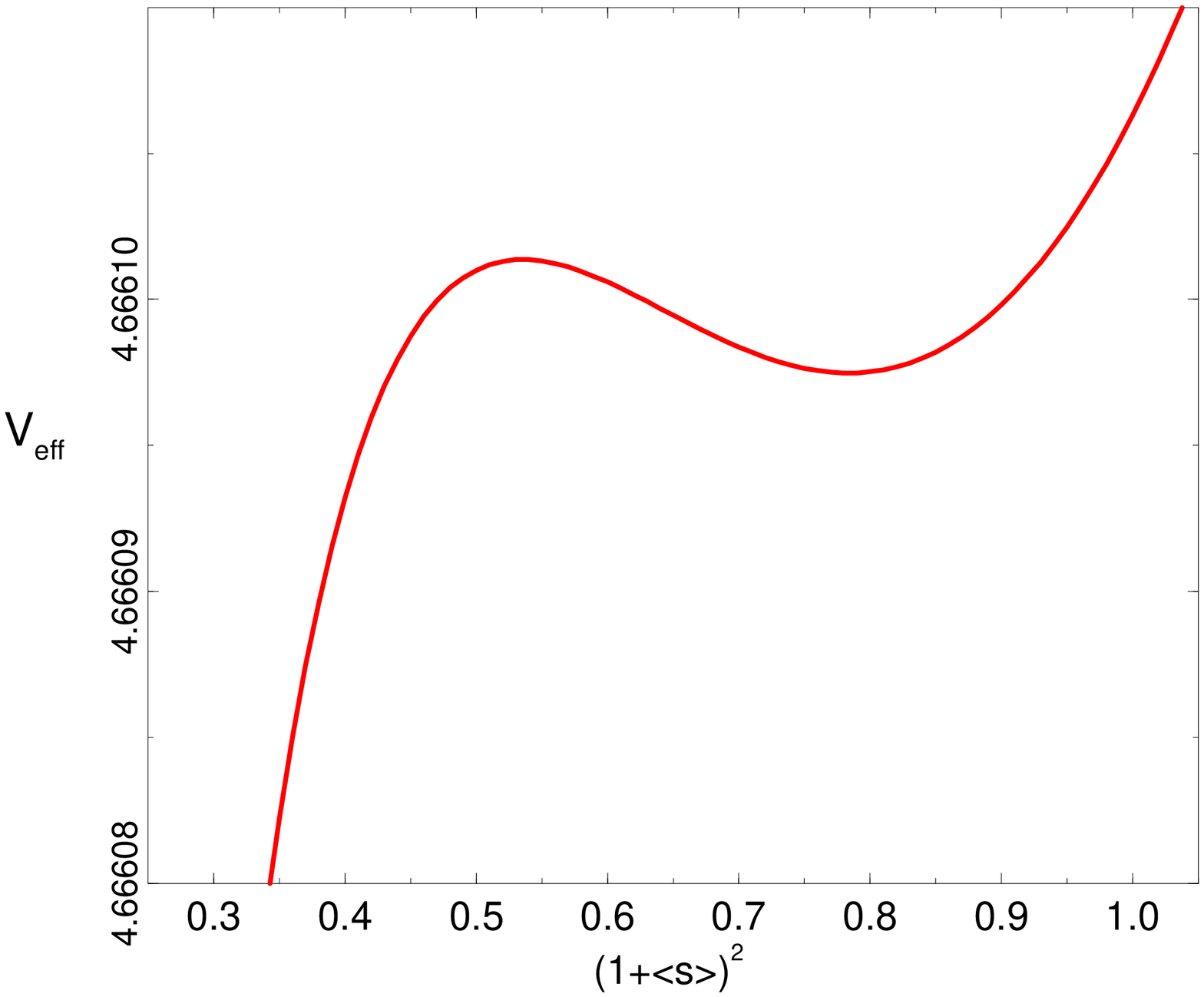}
\caption{The graph of 
$\langle s \rangle$dependence of $V_{eff}$ for $N=200,\bar{g}^2=0.075 
\gtrsim 0.2$.
\\ 
The horizontal axis is $(1+\langle s \rangle)^2$, while the 
vertical axis is $V_{eff}$.}
\label{ng200}
\end{minipage}
\end{figure}

In Fig.~\ref{ng400}, we show the effective potential for $N=400$ 
and $\bar{g}_2=0.075$. We find that there 
is no minimum of the potential at 1-loop level.
Fig.~\ref{ng200} shows the 1-loop potential 
with the same $\bar{g}=g_2a$ but smaller volume $N=200$, where 
we find that there is a minimum. From this fact it becomes clear that 
we cannot take too large volume in the region 
where $\bar{g}$ is not so small. 

\begin{figure}
\includegraphics[width=7cm,clip]{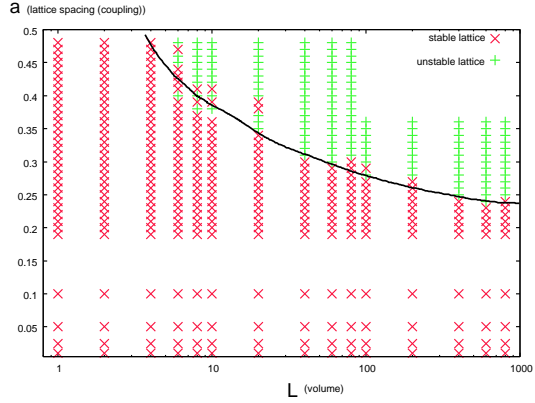}
\caption{
The constraints on the parameter region of ($\bar{g}=g_2a=a,g_2L=L$) 
for the stability of the deconstructed spacetime. 
The symbols $\times$ show the region where the spacetime is stabilized, 
while the symbols $+$ show the region 
where the spacetime is not stabilized.}
\label{OK}
\end{figure}

The above observation suggests that the set of parameters 
($\bar{g}=g_2 a$,  $N=L/a$) or equivalently ($\bar{g}=g_2 a$,  $g_2 L$) 
has to satisfy some constraints in order to stabilize the vacuum 
with a spacetime structure. 
Fig.~\ref{OK} shows the  constraints on 
the parameter region for ($\bar{g}=g_2 a$,  $g_2 L$) 
where 
the deconstructed spacetime can be stabilized. 
We have set $g_2=1$ without loosing generality. 
Lattice spacing $a$ in this graph can be regarded 
as the strength of couplings $\bar{g}$. 
In Fig.~\ref{OK}, the parameter region with stable spacetime structure 
is denoted by the symbol '$\times$' whereas those with no stable 
spacetime structure is denoted by the symbol '$+$'.
It is clear that taking the continuum limit before 
taking the large volume limit 
has a crucial role to stabilize the lattice structure. 

\subsection{Role of the supersymmetry for Deconstruction}\label{boscase}
In the discussion of Sec.~\ref{1pper}, 
it seems that cancellation of 1-loop effect between 
fermion and boson is crucial for stabilizing the 'Deconstruction' vacuum.
In order to see the role of the supersymmetry, we now study the 
\textit{bosonic model} where the fermions are dropped from the 
theory.

The 1-loop effective potential for the bosonic model is 
\begin{align}
V_{eff}(\langle s \rangle)
=&\frac{\bar{\mu}^2}{2\bar{g}^2}[(1+\langle s \rangle)^2-1]^2M\nonumber\\
&+\sum_\mathbf{k}\frac{1}{N^2} 
log[(1+\langle s \rangle)^2
(\widehat{\mathbf{k}}^2+3\bar{\mu}^2)-\bar{\mu}^2]M^2\label{bosptential}
\end{align}
We have to check whether this potential (\ref{bosptential}) 
have stationary point or not. 
The stationary point of this potential $\langle s \rangle_b$ must satisfy 
following equation. 
\begin{align}
0 &= \frac{d}{d\langle s \rangle}V_{eff}(\langle s \rangle_b)\nonumber\\
&=\frac{\bar{\mu}^2}{\bar{g}^2} (1+\langle s \rangle_b)
((1+\langle s \rangle_b)^2-1)M \nonumber\\
&+\frac{1}{N^2}\sum_{\mathbf{k}} 
\frac{2(1+\langle s \rangle_b)
(\hat{\mathbf{k}}^2+3\bar{\mu}^2)}{(1+\langle s
 \rangle_b)^2\hat{\mathbf{k}}^2
 +
(3(1+\langle s \rangle_b)^2-1)\bar{\mu}^2}M^2. 
\label{needboson}
\end{align}
In order to have a well-defined perturbative vacuum with no tachyons
$\frac{1}{3}<(1+\langle s \rangle)^2$ has to be satisfied. 
In this region the second term in Eq.~(\ref{needboson}) is positive.
Therefore, the minimum can only exist in the region 
$(1+\langle s \rangle)^2 < 1$, since otherwise the first term 
in Eq.~(\ref{needboson}) is 
also positive. Now when $\frac{1}{3}<(1+\langle s \rangle)^2 < 1$, 
the second term is larger than $M^2$, and first term is larger than 
$-\frac{\bar{\mu}^2}{\bar{g}^2}M$ so that 
\begin{equation}
\frac{d}{d\langle s  \rangle}V_{eff} > M^2-\frac{\bar{\mu}^2}{\bar{g}^2}M.
\end{equation}
Since $M>1$ for non-abelian gauge group, 
\begin{eqnarray}
\frac{d}{d\langle   s \rangle }V_{eff} > 0,&\mbox{when}&
\frac{\bar{\mu}}{\bar{g}}=(g_2L)^{-1} < 1,
\label{dV} 
\end{eqnarray}
so that there is no stable minimum near the tree level minimum 
in all the physically natural parameter region as given in 
Eq.~(\ref{physreg1}).
Although it is difficult to find the global minimum with large order 
correction in perturbation theory, Eq.~(\ref{dV}) suggests that 
the quantum effect in the bosonic model has the effect 
to drive VEV $(1+\langle s \rangle)/\sqrt{2}a$ 
to smaller value, which corresponds to larger lattice spacing.  
Eq.~(\ref{dV}) imply 
that  
\textit{the deconstruction cannot make a stable spacetime 
if there is no 
fermion-boson cancellation.}
The instability of the bosonic model was also observed 
by Giedt~\cite{Giedt 0312} in his non-perturbative 
study 
 on the
bosonic part of the CKKU model for the (4,4) 2d
super-Yang-Mills~\cite{Kaplan3}
where he found that the bosonic fields $x$ 
seem to concentrate at $\langle x \rangle \sim 0$.
From these observation one can say the CKKU model has a stable 
vacuum not simply because the quantum correction is small but 
because the quantum corrections from the fermionic modes and bosonic modes 
cancel, 
which means that the supersymmetry 
is crucial for the stabilization of the deconstruction. 

\subsection{The meaning of 1-loop calculation for the study of 
stabilization of lattice spacing}
We calculated the 1-loop effective potential using the Coleman-Weinberg 
method. In order to study the stability of the theory, of course one has 
to carry out non-perturbative analyses eventually. However, it would be 
still useful to study the effective potential in 1-loop approximation for 
two reasons;
(1) By obtaining analytical forms of the potential or correlation 
    functions at 1-loop, one can understand the detailed structure of 
    quantum corrections. 
(2) The perturbative  result would also be useful for future non-perturbative 
    numerical calculations, since it gives us a quantitative idea on the 
    appropriate parameter region in which the simulation should be carried 
    out with good stability of the vacuum and scaling property.

\section{Conclusion and discussion}\label{Sec:summary}

 In this paper, we have studied the CKKU model at the quantum level 
by an explicit perturbative calculation for the case of $U(2)$
gauge group. 
We have pointed the subtleties of the perturbative correction in CKKU 
model which arises from the zero-eigenvalue of the fermion-matrix 
and the massless zero momentum mode of the bosonic fields.
To make the fermion path-integral well-defined, we have introduced 
the fermion mass term in Eq.~(\ref{fermimass}), although we have find that 
the two-step limit,  where we take the limit of zero fermion mass at first 
before taking the continuum limit, is necessary to control the 
counter terms. In order to avoid the infra-red divergences
we have separated the zero momentum modes and carried out the path-integral 
over these fields non-perturbatively, 
while non-zero momentum  modes are treated 
perturbatively.

We have then studied the possible counter terms in this model, namely 
the bosonic 1-point and 2-point functions by the explicit calculation.
We have found that there are non-trivial quantum mass corrections 
larger than the tree level mass. However these corrections 
vanish in the infinite volume limit so that the CKKU model does not 
need fine-tuning to recover the full supersymmetry.

We have also studied the stability of the 
lattice spacetime generated by the ^^ deconstruction'. 
We have found the constraint on the parameter region 
for which the lattice spacetime is stable against 1-loop 
corrections. 
This constraint is not only interesting 
quantitative information of the property of the lattice spacetime, 
but also practically useful as a guide for the fully numerical 
non-perturbative simulations in the future. 
We have also understood that the cancellation of quantum corrections 
between bosons and fermions 
is  
crucial to stabilize the lattice spacetime.

It would of course be important to make a full non-perturbative 
study of CKKU model in our prescription.  In particular, 
a comparison of the result with the study by Giedt~\cite{Giedt 0405} 
by a phase quenched model would be interesting.
Recently, there are new lattice theories which preserve the supersymmetry 
on the lattice
\cite{Catterall:2001wx}-\cite{Kanamori2}.
These are the models without 
deconstruction, and might be useful for the practical 
study. The perturbative study of these models would 
also be important.

\section*{Acknowledgments}

The authors would like to thank  T. Umeda, H. Fukaya, T. Goto,
H. Shimada, and
S. Sugimoto for stimulating discussions. 
We would like to
give special thanks to
H. Fukaya and H. Shimada 
for reading the manuscript and 
giving crucial comments.
We also acknowledge the
Yukawa Institute for Theoretical Physics at Kyoto 
University, where this work was benefited from the discussions during 
the YITP-W-04-19 workshop on "New Trends in Lattice Field Theory''. 
T.~O. is supported by Grant-in-Aid for Scientific research 
from the Ministry of Education, Culture, Sports, Science and
Technology of Japan (Nos. 13135213,16028210, 16540243). 

\appendix
\section{Notation and fourier transformation}\label{Sec:fourier}
Now we define the fourier transformation 
of the fields on the lattice 
as follows.
\begin{align}
&\tilde{s}_x(\mathbf{0}) = a\frac{1}{\sqrt{\bar{g}N^3}}
\sum_{\mathbf{n}} 
s_{x\mathbf{n}}
\quad 
\tilde{v}_x(\mathbf{0})=
\sum_{\mathbf{n}} 
a\frac{1}{\sqrt{\bar{g}N^3}}v_{x\mathbf{n}} 
\label{cano1} \\
&\tilde{s}_y(\mathbf{0})=
\sum_{\mathbf{n}} 
a\frac{1}{\sqrt{\bar{g}N^3}} s_{y\mathbf{n}}
\quad 
\tilde{v}_y(\mathbf{0})=
\sum_{\mathbf{n}} a\frac{1}{\sqrt{\bar{g}N^3}}v_{y\mathbf{n}}
\label{cano2}
\end{align}
\begin{align}
&\tilde{s}_x(\mathbf{k}\ne \mathbf{0}) = a\frac{1}{\bar{g}N}
\sum_{\mathbf{n}} 
s_{x\mathbf{n}}e^{-ia\mathbf{k} \cdot \mathbf{n}}e^{-ia\frac{1}{2}k_x}
\quad 
\tilde{v}_x(\mathbf{k}\ne \mathbf{0})=
\sum_{\mathbf{n}} 
a\frac{1}{\bar{g}N}v_{x\mathbf{n}}
e^{-ia\mathbf{k} \cdot \mathbf{n}}e^{-ia\frac{1}{2}k_x}\label{cano1a} \\
&\tilde{s}_y(\mathbf{k}\ne \mathbf{0})=
\sum_{\mathbf{n}} 
a\frac{1}{\bar{g}N} s_{y\mathbf{n}} e^{-ia\mathbf{k} \cdot \mathbf{n}}e^{-ia\frac{1}{2}k_y}
\quad 
\tilde{v}_y(\mathbf{k}\ne \mathbf{0})=
\sum_{\mathbf{n}} a\frac{1}{\bar{g}N} v_{y\mathbf{n}}
e^{-ia\mathbf{k} \cdot \mathbf{n}}e^{-ia\frac{1}{2}k_y} \label{cano2a}
\end{align}
\begin{eqnarray}
&\hat{x}_{\mathbf{0}}=a\frac{1}{\sqrt{\bar{g}N^3}}\sum_{\mathbf{n}} 
(x_{\mathbf{n}}-\frac{1}{\sqrt{2}a})
\quad 
\hat{\bar{x}}_{\mathbf{0}}=a\frac{1}{\sqrt{\bar{g}N^3}}\sum_{\mathbf{n}} 
(\bar{x}_{\mathbf{n}}-\frac{1}{\sqrt{2}a})
\nonumber\\
&\hat{y}_{\mathbf{0}}=a\frac{1}{\sqrt{\bar{g}N^3}}\sum_{\mathbf{n}} 
(y_{\mathbf{n}}-\frac{1}{\sqrt{2}a})
\quad 
\hat{\bar{y}}_{\mathbf{0}}=a\frac{1}{\sqrt{\bar{g}N^3}}\sum_{\mathbf{n}} 
(\bar{y}_{\mathbf{n}}-\frac{1}{\sqrt{2}a})
\end{eqnarray}
\begin{eqnarray}
&\hat{x}_{\mathbf{k}\ne 0}=a\frac{1}{\bar{g}N}\sum_{\mathbf{n}} 
(x_{\mathbf{n}}-\frac{1}{\sqrt{2}a})e^{-ia\mathbf{k}\cdot\mathbf{n}}
\quad 
\hat{\bar{x}}_{\mathbf{k}\ne 0}=a\frac{1}{\bar{g}N}\sum_{\mathbf{n}} 
(\bar{x}_{\mathbf{n}}-\frac{1}{\sqrt{2}a})e^{-ia\mathbf{k}\cdot\mathbf{n}}
\nonumber\\
&\hat{y}_{\mathbf{k}\ne 0}=a\frac{1}{\bar{g}N}\sum_{\mathbf{n}} 
(y_{\mathbf{n}}-\frac{1}{\sqrt{2}a})e^{-ia\mathbf{k}\cdot\mathbf{n}}
\quad 
\hat{\bar{y}}_{\mathbf{k}\ne 0}=a\frac{1}{\bar{g}N}\sum_{\mathbf{n}} 
(\bar{y}_{\mathbf{n}}-\frac{1}{\sqrt{2}a})e^{-ia\mathbf{k}\cdot\mathbf{n}}
\end{eqnarray}

\begin{align}
&\tilde{\alpha}_{\mathbf{0}} = a^\frac{3}{2}\frac{1}{\sqrt{\bar{g}N^3}}
\sum_{\mathbf{n}} 
\alpha_\mathbf{n}e^{-ia\mathbf{k} \cdot \mathbf{n}}e^{-ia\frac{1}{2}k_x}
\quad 
\tilde{\beta}_{\mathbf{0}}=a^\frac{3}{2}\frac{1}{\sqrt{\bar{g}N^3}}
\sum_{\mathbf{n}} 
 \beta_\mathbf{n} e^{-ia\mathbf{k} \cdot \mathbf{n}}e^{-ia\frac{1}{2}k_y} 
\label{cano3} \\
&\tilde{\lambda}_{\mathbf{0}}=a^\frac{3}{2}\frac{1}{\sqrt{\bar{g}N^3}}
\sum_{\mathbf{n}} 
\lambda_\mathbf{n} e^{-ia\mathbf{k} \cdot \mathbf{n}}
\quad 
\tilde{\xi}_{\mathbf{0}}=a^\frac{3}{2}\frac{1}{\sqrt{\bar{g}N^3}}
\sum_{\mathbf{n}}  \xi_\mathbf{n}e^{-ia\mathbf{k} \cdot \mathbf{n}}e^{-ia\frac{1}{2}(k_x+k_y)}\label{cano4}
\end{align}
\begin{align}
&\tilde{\alpha}_{\mathbf{k}\ne \mathbf{0}} = a^\frac{3}{2}\frac{1}{\bar{g}N}
\sum_{\mathbf{n}} 
\alpha_\mathbf{n}e^{-ia\mathbf{k} \cdot \mathbf{n}}e^{-ia\frac{1}{2}k_x}
\quad 
\tilde{\beta}_{\mathbf{k}\ne \mathbf{0}}=a^\frac{3}{2}\frac{1}{\bar{g}N}
\sum_{\mathbf{n}} 
 \beta_\mathbf{n} e^{-ia\mathbf{k} \cdot \mathbf{n}}e^{-ia\frac{1}{2}k_y} 
\label{cano3a} \\
&\tilde{\lambda}_{\mathbf{k}\ne \mathbf{0}}=a^\frac{3}{2}\frac{1}{\bar{g}N}
\sum_{\mathbf{n}} 
\lambda_\mathbf{n} e^{-ia\mathbf{k} \cdot \mathbf{n}}
\quad 
\tilde{\xi}_{\mathbf{k}\ne \mathbf{0}}=a^\frac{3}{2}\frac{1}{\bar{g}N}
\sum_{\mathbf{n}}  \xi_\mathbf{n}e^{-ia\mathbf{k} \cdot \mathbf{n}}e^{-ia\frac{1}{2}(k_x+k_y)}\label{cano4a}
\end{align}

We denote two-dimensional momentum as $\mathbf{k}\equiv (k_x,k_y)$. 
And we define the lattice momentum $\widehat{k_{x,y}},
\widehat{\mathbf{k}}$ 
as 
\begin{align}
&\widehat{k'_x}=\frac{2}{a}\sin (\frac{ak_x}{2}),\quad 
\widehat{k_x}=a\widehat{k'_x}= 2\sin (\frac{ak_x}{2})\\
&\widehat{k'_y}=\frac{2}{a}\sin (\frac{ak_y}{2}),\quad 
\widehat{k_y}=a\widehat{k'_y}= 2\sin (\frac{ak_y}{2})\\
&\widehat{\mathbf{k}}^2\equiv\widehat{k_x}^2+\widehat{k_y}^2
\quad 
\end{align}
The coupling $\bar{g}$, 
and the masses $\bar{\mu},\bar{\mu_F}$
in the lattice unit
are defined as
\begin{equation}
\bar{g}=g_2a, \bar{\mu}=a\mu=\frac{1}{N},\bar{\mu_F}=a\mu_F
\end{equation}
\\

We denote generators of gauge group 
$U(2)$ with fundamental representation as 
$2 \times 2$ matrices $T^{\mu},(\mu=0,1,2,3)$ 
,where $T^0$ is the one of $U(1) \subset U(2)$ and 
$T^1,T^2,T^3$ are ones for $SU(2) \subset U(2)$.

We also define $t^{\mu\nu\rho}$ and
$t^{\mu\nu\rho}_{\mathbf{m},\mathbf{n}}$ 
as
\begin{align}
&t^{\mu\nu\rho}_{\mathbf{m},\mathbf{n}}
\equiv t^{\mu\nu\rho} \delta_{\mathbf{m},\mathbf{n}}
\equiv \frac{1}{2}Tr(T^{\mu}T^{\nu}T^{\rho})\delta_{\mathbf{m},\mathbf{n}}.
\end{align}

\section{fermion matrix}\label{Sec:fermionmatrix}

Momentum representation of fermion action 
is written as 
\begin{align}
&\bar{g}^{-2}N^2
\Bigl(
\begin{array}{cc|cc|c|c}
\tilde{\alpha}_{\mathbf{k}}^\mu,&\tilde{\beta}_{\mathbf{k}}^\mu,&
\tilde{\alpha}_{\mathbf{0}}^{a(\ne 0)},&\tilde{\beta}_{\mathbf{0}}^{a(\ne 0)},&
\tilde{\beta}_{\mathbf{0}}^0,&\tilde{\alpha}_{\mathbf{0}}^0
\end{array}
\Bigr)\cdot \nonumber\\
&\left( \begin{array}{c|c|c|c}
\bar{g}^2\bar{A}_{(2,2)\mathbf{k},\mathbf{p}}^{\mu\nu}
& \bar{g}^{\frac{5}{2}} B_{(2,2)\mathbf{k},\mathbf{0}}^{\mu b}
& \bar{g}^{\frac{5}{2}} C_{\xi (2,1)\mathbf{k},\mathbf{0}}^{\mu 0}
& \bar{g}^{\frac{5}{2}}\bar{\mu}_F 
C_{\lambda (2,1)\mathbf{k},\mathbf{0}}^{\mu 0}
\\ \hline
\bar{g}^{\frac{5}{2}} D_{(2,2)\mathbf{0},\mathbf{p}}^{a\nu}
& \bar{g}\bar{\mu}_F E'_{2,2}+
\bar{g}^{\frac{3}{2}} E_{(2,2)\mathbf{0},\mathbf{0}}^{ab}
& \bar{g}^{\frac{3}{2}}\bar{\mu}_F F_{\xi (2,1)\mathbf{0},\mathbf{0}}^{a0}& \bar{g}^{\frac{3}{2}}\bar{\mu}_F 
F_{\lambda (2,1)\mathbf{0},\mathbf{0}}^{a0} \\ \hline 
\bar{g}^{\frac{5}{2}} G_{(1,2)\mathbf{0},\mathbf{p}}^{0\nu}&
\bar{g}^{\frac{3}{2}}\bar{\mu}_F H_{\lambda (2,1)\mathbf{0},\mathbf{0}}^{0b} & \bar{g}\bar{\mu}_F & -\bar{g}\bar{\mu}_F \\ \hline 
\bar{g}^{\frac{5}{2}}J_{(1,2)\mathbf{0},\mathbf{p}}^{0\nu}
& \bar{g}^{\frac{3}{2}}\bar{\mu}_FK_{(1,2)\mathbf{0},\mathbf{0}}^{0b} 
& \bar{g}\bar{\mu}_F 
& \bar{g}\bar{\mu}_F 
\end{array}
\right)
\left(
\begin{array}{c}
\tilde{\lambda}_{\mathbf{p}}^\nu \\
\tilde{\xi}_{\mathbf{p}}^\nu \\
\hline
\tilde{\lambda}_{\mathbf{0}}^{a(\ne 0)} \\
\tilde{\xi}_{\mathbf{0}}^{a(\ne 0)} \\
\hline
\tilde{\xi}_{\mathbf{0}}^0 \\
\hline
\tilde{\lambda}_{\mathbf{0}}^0
\end{array}
\right)
\label{momfer2},
\end{align}
where the sub-matrices
$\bar{A}_{(2,2)\mathbf{k},\mathbf{p}}^{\mu\nu},
B_{(2,2)\mathbf{k},\mathbf{0}}^{\mu b}, \cdots 
,K_{(1,2)\mathbf{0},\mathbf{0}}^{0b} $ are of order unity with respect to 
$\bar{\mu}$,$\bar{\mu}_{F}$ and $\bar{g}$.
In this section, we take generators with subscripts
written in Roman letters $a,b$ as the 
generators of $SU(2) \subset U(2)$, and ones with subscripts written in 
Greek characters $\mu,\nu$ as generators of $U(2)$.
The explicit forms of sub-matrices of fermion matrix are 
given as follows:
\begin{align}
&\bar{A}_{(2,2)\mathbf{k},\mathbf{p}}^{\mu\nu}
= \left( 
\begin{array}{cc}
a_{1,1} & a_{1,2} \\
a_{2,1} & a_{2,2} 
\end{array}
\right), 
\end{align}
where 
\begin{align}
a_{1,1} =&(i\widehat{k_x}+\bar{\mu}_F e^{(\frac{iak_x}{2})})(1+\langle s \rangle)
\delta_{\mathbf{k}+\mathbf{p},\mathbf{0}}\delta^{\mu\nu} \nonumber\\
&+(\bar{g}\hat{\bar{x}}_{\mathbf{q}\ne \mathbf{0}}+
\bar{g}^{\frac{1}{2}}\hat{\bar{x}}_{\mathbf{q} = \mathbf{0}}
)^\rho 
(t^{\mu\rho\nu}e^{(\frac{-ip_x}{2})} (1+\bar{\mu}_F)-t^{\mu\nu\rho}e^{(\frac{ip_x}{2})}) 
\delta_{\mathbf{k}+\mathbf{p}+\mathbf{q},\mathbf{0}},\\
a_{1,2} =&(i\widehat{k_y}-\bar{\mu}_F e^{(\frac{-iak_y}{2})})(1+\langle s \rangle)
 \delta_{\mathbf{k}+\mathbf{p},\mathbf{0}}\delta^{\mu\nu} \nonumber\\
&+(\bar{g}\hat{y}_{\mathbf{q}\ne \mathbf{0}}+
\bar{g}^{\frac{1}{2}}\hat{y}_{\mathbf{q} = \mathbf{0}}
)^\rho
(-t^{\mu\rho\nu}e^{(\frac{i(q_x-k_y)}{2})}(1+\bar{\mu}_F)
+t^{\mu\nu\rho}e^{(\frac{-i(q_x-k_y)}{2})})
\delta_{\mathbf{k}+\mathbf{p}+\mathbf{q},\mathbf{0}}, \\
a_{2,1} =&(i\widehat{k_y}+\bar{\mu_F} e^{(\frac{iak_y}{2})})(1+\langle s \rangle)
\delta_{\mathbf{k}+\mathbf{p},\mathbf{0}}\delta^{\mu\nu} \nonumber\\
+&(\bar{g}\hat{\bar{y}}_{\mathbf{q}\ne \mathbf{0}}+
\bar{g}^{\frac{1}{2}}\hat{\bar{y}}_{\mathbf{q} = \mathbf{0}}
)^\rho
(t^{\mu\rho\nu}e^{(\frac{-iap_x}{2})}(1+\bar{\mu}_F)-t^{\mu\nu\rho}e^{(\frac{iap_y}{2})})
\delta_{\mathbf{k}+\mathbf{p}+\mathbf{q},\mathbf{0}}, \\
a_{2,2} =&(-i\widehat{k_x}+\bar{\mu_F} e^{(\frac{-iak_x}{2})})(1+\langle s \rangle)
\delta_{\mathbf{k}+\mathbf{p},\mathbf{0}}\delta^{\mu\nu} \nonumber\\
&+(\bar{g}\hat{x}_{\mathbf{q}\ne \mathbf{0}}+
\bar{g}^{\frac{1}{2}}\hat{x}_{\mathbf{q} = \mathbf{0}}
)^\rho 
(t^{\mu\rho\nu}e^{(\frac{ia(q_y-k_x)}{2})}(1+\bar{\mu}_F) 
-t^{\mu\nu\rho}e^{(\frac{-ia(q_y-k_x)}{2})}))
\delta_{\mathbf{k}+\mathbf{p}+\mathbf{q},\mathbf{0}}. 
\label{A2}
\end{align}

\begin{equation}
E_{(2,2)\mathbf{0},\mathbf{0}}^{ab}
= \left( \begin{array}{c|c}
\hat{\bar{x}}_{\mathbf{0}}^\rho 
(t^{a \rho b}(1+\bar{\mu_F})-t^{ab \rho}) 
&\hat{y}_{\mathbf{0}}^\rho 
(-t^{a\rho b}(1+\bar{\mu_F})+t^{ab \rho}) 
\\
\hline
\hat{\bar{y}}_{\mathbf{0}}^\rho 
(t^{a\rho b}(1+\bar{\mu_F})-t^{ab \rho}) 
&\hat{x}_{\mathbf{0}}^\rho 
(t^{a\rho b}(1+\bar{\mu_F})-t^{ab\rho}) 
\end{array}
\right)\label{E}
\end{equation}
\begin{equation}
\bar{\mu_F} E'_{2,2} 
= \left( \begin{array}{cc}
\bar{\mu}_F \delta^{ab} (1+\langle s \rangle)&
-\bar{\mu}_F \delta^{ab} (1+\langle s \rangle)\\
\bar{\mu}_F \delta^{ab} (1+\langle s \rangle)&
\bar{\mu}_F \delta^{ab}(1+\langle s \rangle)
\end{array}
\right)\label{Em}
\end{equation}
\begin{equation}
B_{(2,2)\mathbf{k},\mathbf{0}}^{\mu b}= \left( 
\begin{array}{c|c}
\hat{\bar{x}}_{-\mathbf{k}}^\rho 
(t^{\mu\rho b}(1+\bar{\mu}_F)-t^{\mu b \rho})
& \hat{y}_{-\mathbf{k}}^\rho 
(-t^{\mu\rho b}e^{(\frac{-ia(k_y+k_x)}{2})}(1+\bar{\mu}_F)
+t^{\mu b \rho}e^{(\frac{ia(k_y+k_x)}{2})}) 
\\
\hline
\hat{\bar{y}}_{-\mathbf{k}}^\rho 
(t^{\mu\rho b }(1+\bar{\mu}_F)-t^{\mu b \rho}) 
& \hat{x}_{-\mathbf{k}}^\rho 
(t^{\mu\rho b}e^{(\frac{-ia(k_y+k_x)}{2})}(1+\bar{\mu}_F)
-t^{\mu b \rho}e^{(\frac{ia(k_y+k_x)}{2})})
\end{array}
\right)
\end{equation}

\begin{equation}
C_{\xi(2,1)\mathbf{k},\mathbf{0}}^{\mu 0}= \left( 
\begin{array}{c}
\hat{y}_{-\mathbf{k}}^\rho 
(-t^{\mu\rho 0}e^{(\frac{-ia(k_y+k_x)}{2})}(1+\bar{\mu}_F)
+t^{\mu 0 \rho}e^{(\frac{ia(k_y+k_x)}{2})}) 
\\
\hat{x}_{-\mathbf{k}}^\rho 
(t^{\mu\rho 0}e^{(\frac{-ia(k_y+k_x)}{2})}(1+\bar{\mu}_F)
-t^{\mu 0 \rho}e^{(\frac{ia(k_y+k_x)}{2})})
\end{array}
\right)
\quad
C_{\lambda (2,1)\mathbf{k},\mathbf{0}}^{\mu 0}= \left( 
\begin{array}{c}
\hat{\bar{x}}_{-\mathbf{k}}^\rho 
(t^{\mu \rho 0})
\\
\hat{\bar{y}}_{-\mathbf{k}}^\rho 
(t^{\mu\rho 0}) 
\end{array}
\right)
\end{equation}

\begin{equation}
D_{(2,2)\mathbf{0},\mathbf{p}}^{a \nu}= \left( 
\begin{array}{c|c}
\hat{\bar{x}}_{-\mathbf{p}}^\rho 
(t^{a \rho \nu}e^{(\frac{-ip_x}{2})}(1+\bar{\mu}_F)
-t^{a \nu \rho}e^{(\frac{ip_x}{2})})
& \hat{y}_{-\mathbf{p}}^\rho 
(-t^{a \rho \nu}e^{(\frac{-ip_x}{2})}(1+\bar{\mu}_F)
+t^{a \nu \rho}e^{(\frac{ip_x}{2})})
\\
\hline
\hat{\bar{y}}_{-\mathbf{p}}^\rho 
(t^{a \rho \nu}e^{(\frac{-ip_y}{2})}(1+\bar{\mu}_F)
-t^{a \nu \rho}e^{(\frac{ip_y}{2})})
& \hat{x}_{-\mathbf{p}}^\rho 
(t^{a \rho \nu}e^{(\frac{-ip_y}{2})}(1+\bar{\mu}_F)
-t^{a \nu \rho}e^{(\frac{ip_y}{2})})
\end{array}
\right)
\end{equation}

\begin{equation}
F_{\xi(2,1)\mathbf{0},\mathbf{0}}^{a 0}= \left( 
\begin{array}{c}
-\hat{y}_{\mathbf{0}}^a 
(t^{a a 0}) 
\\
\hat{x}_{\mathbf{0}}^a 
(t^{a a 0}) 
\end{array}
\right)
\quad
F_{\lambda (2,1)\mathbf{0},\mathbf{0}}^{a 0}= \left( 
\begin{array}{c}
\hat{\bar{x}}_{\mathbf{0}}^a 
t^{a a 0}
\\
\hat{\bar{y}}_{\mathbf{0}}^a 
(t^{a a 0}) 
\end{array}
\right)
\end{equation}

\begin{equation}
G_{(1,2)\mathbf{0},\mathbf{p}}^{0 \nu}= \left( 
\hat{\bar{x}}_{-\mathbf{p}}^\rho 
(t^{0 \rho \nu}e^{(\frac{-ip_x}{2})}(1+\bar{\mu}_F)
-t^{0 \nu \rho}e^{(\frac{ip_x}{2})})
,\hat{y}_{-\mathbf{p}}^\rho 
(-t^{0 \rho \nu}e^{(\frac{-ip_x}{2})}(1+\bar{\mu}_F)
+t^{0 \nu \rho}e^{(\frac{ip_x}{2})})
\right)
\end{equation}

\begin{equation}
H_{(1,2)\mathbf{0},\mathbf{0}}^{0 b}= 
\left( 
\hat{\bar{x}}_{\mathbf{0}}^\rho 
(t^{0 \rho b}(1+\bar{\mu_F})-t^{0b \rho}) 
,\hat{y}_{\mathbf{0}}^\rho 
(-t^{0\rho b}(1+\bar{\mu_F})+t^{0b \rho}) 
\right)
\end{equation}

\begin{equation}
J_{(1,2)\mathbf{0},\mathbf{p}}^{0 \nu}= \left( 
\hat{\bar{y}}_{-\mathbf{p}}^\rho 
(t^{0 \rho \nu}e^{(\frac{-ip_y}{2})}(1+\bar{\mu}_F)
-t^{0 \nu \rho}e^{(\frac{ip_y}{2})})
,\hat{x}_{-\mathbf{p}}^\rho 
(t^{0 \rho \nu}e^{(\frac{-ip_y}{2})}(1+\bar{\mu}_F)
-t^{0 \nu \rho}e^{(\frac{ip_y}{2})}) \right)
\end{equation}

\begin{equation}
K_{(1,2)\mathbf{0},\mathbf{0}}^{0 b}= \left( 
\hat{\bar{y}}_{\mathbf{0}}^\rho 
(t^{0\rho b}(1+\bar{\mu_F})-t^{0b \rho}) 
,\hat{x}_{\mathbf{0}}^\rho 
(t^{0\rho b}(1+\bar{\mu_F})-t^{0b\rho}) 
\right)
\end{equation}

The ghost term is described as
\begin{eqnarray}
S_{gh} 
&=& \bar{g}^{-1}N
\Bigl( (\frac{\bar{g}}{N})^{1/2} \bar{\tilde{c}}_{\mathbf{k}}^\mu, 
  \tilde{\bar{c}}_{\mathbf{0}}^a, 
  \tilde{\bar{c}}_{\mathbf{0}}^0 
\Bigr)
\left( \begin{array}{c|c|c}
  \bar{\Upsilon}_{\mathbf{k},\mathbf{p}}^{\mu\nu}
& \left( \frac{\bar{g}}{N} \right) \Theta_{\mathbf{k},\mathbf{0}}^{\mu b}
& 0 
\\ \hline
  0 & 0 & 0 \\ \hline 
  0 & 0 & 0 
\end{array}
\right)
\left(
\begin{array}{c}
(\frac{\bar{g}}{N})^{1/2}\tilde{c}_{\mathbf{p}}^\nu \\
 \tilde{c}_{\mathbf{0}}^a \\
 \tilde{c}_{\mathbf{0}}^0 
\end{array}
\right),
\label{ghmatapp:A}
\end{eqnarray}
where 
\begin{align}
\bar{\Upsilon}_{\mathbf{k},\mathbf{p}}^{\mu\nu}=
&\widehat{\mathbf{k}}^2\delta^{\mu,\nu}\delta_{\mathbf{k},\mathbf{p}}
-(e^{ik_x}-1)[iTr(T^{\mu}[T^{\nu},\hat{x}_{-\mathbf{k}-\mathbf{p}}])
-iTr (T^{\mu}\hat{x}_{-\mathbf{k}-\mathbf{p}}T^{\nu})(e^{ip_x}-1)\nonumber\\
&-iTr(T^{\mu}[\hat{\bar{x}}_{-\mathbf{k}-\mathbf{p}},T^{\nu}])
-iTr (T^{\mu}T^{\nu}\hat{\bar{x}}_{-\mathbf{k}-\mathbf{p}})(e^{ip_x}-1)]
\nonumber\\
&-(e^{ik_y}-1)[iTr(T^{\mu}[T^{\nu},\hat{y}_{-\mathbf{k}-\mathbf{p}}])
-iTr (T^{\mu}\hat{y}_{-\mathbf{k}-\mathbf{p}}T^{\nu})(e^{ip_y}-1)\nonumber\\
&-iTr(T^{\mu}[\hat{\bar{y}}_{-\mathbf{k}-\mathbf{p}},T^{\nu}])
-iTr (T^{\mu}T^{\nu}\hat{\bar{y}}_{-\mathbf{k}-\mathbf{p}})(e^{ip_y}-1)],
\\
 \Theta_{\mathbf{k},\mathbf{0}}^{\mu b}=&
-(e^{ik_x}-1)[iTr(T^{\mu}[T^{b},\hat{x}_{-\mathbf{k}}])
-iTr(T^{\mu}[\hat{\bar{x}}_{-\mathbf{k}},T^{b}])]\nonumber\\
&-(e^{ik_y}-1)[iTr(T^{\mu}[T^{b},\hat{y}_{-\mathbf{k}}])
-iTr(T^{\mu}[\hat{\bar{y}}_{-\mathbf{k}},T^{b}])]
\end{align}

\section{Measure term}\label{Sec:measure}
Here, we will give the expression of the measure term. 
The gauge invariant measure term $\sqrt{det(g)_{\mathbf{n}}}$ 
is defined by the metric $g_{AB\mathbf{n}}$ as 
\begin{equation}
\int d\bar{x}dx d\bar{y}dy 
=\int \displaystyle{\prod_{\mathbf{n}}}\sqrt{det(g)}ds_xds_ydv_xdv_y,
\end{equation}
where the metric is defined by the gauge invariant norm 
\begin{equation}
Tr[(dx_{\mathbf{n}}d\bar{x}_{\mathbf{n}})
+(dy_{\mathbf{n}}d\bar{y}_{\mathbf{n}})]
=g_{AB\mathbf{n}} 
d\phi^A_{\mathbf{n}} d\phi^B_{\mathbf{n}} . \label{metric}
\end{equation}
Here $\phi^{A}_{\mathbf{n}} (A=1, \cdots , 16)$ represents the $U(2)$ 
scalar and vector fields 
namely $\{s^{\alpha}_{x\mathbf{n}},v^{\alpha}_{x\mathbf{n}},
s^{\alpha}_{y\mathbf{n}},v^{\alpha}_{y\mathbf{n}} ;\alpha=0,1,2,3\}$
Using the parameterizations in Eqs.~(\ref{y1}) and (\ref{bar1}),
$dx_{\mathbf{n}}$ is written as 
\begin{align}
dx_{\mathbf{n}}=ds_{x\mathbf{n}}U_{x\mathbf{n}}+(1+\langle s \rangle+s_{x\mathbf{n}})dU_{x\mathbf{n}} \quad dU_{x\mathbf{n}}=e^{ia(v_{x\mathbf{n}}+dv_{x\mathbf{n}})}-e^{iav_{x\mathbf{n}}}  \\
d\bar{x}_{\mathbf{n}}=U^{\dagger}_{x\mathbf{n}}ds_{x\mathbf{n}}+dU^{\dagger}_{x\mathbf{n}}(1+\langle s \rangle+s_{x\mathbf{n}})
\quad dU^{\dagger}_{x\mathbf{n}}=e^{-ia(v_{x\mathbf{n}}+dv_{x\mathbf{n}})}-e^{-iav_{x\mathbf{n}}}. 
\end{align}
Then, the left hand side of (\ref{metric}) will be
\begin{align}
&Tr( dx_\mathbf{n}d\bar{x}_\mathbf{n})\nonumber\\
&=\frac{1}{2}Tr[ds_{x\mathbf{n}}^2
+ds_{x\mathbf{n}}(U_{x\mathbf{n}}dU^{\dagger}_{x\mathbf{n}})z_{x\mathbf{n}}
+z_{x\mathbf{n}}(dU_{x\mathbf{n}}U^{\dagger}_{x\mathbf{n}})ds_{x\mathbf{n}}
+z_{x\mathbf{n}}^2dU_{x\mathbf{n}}dU^{\dagger}_{x\mathbf{n}}],
\label{norm}
\end{align}
where
\begin{equation}
z_{x\mathbf{n}}=(1+\langle s \rangle+s_{x\mathbf{n}}).
\end{equation}
Explicit form of $dU_{x\mathbf{n}}U^{\dagger}_{x\mathbf{n}}$ 
is obtained as 
\begin{align}
U_{x\mathbf{n}}dU^{\dagger}_{x\mathbf{n}}
=-i\frac{e^{i\bar{V}_{x\mathbf{n}}}-1}{i\bar{V}_{x\mathbf{n}}} dv_{x\mathbf{n}}
=-iT^{\alpha} 
\left( \frac{e^{i\bar{V}_{x\mathbf{n}}}-1}{i\bar{V}_{x\mathbf{n}}} 
\right)_{\alpha\beta} dv_{x\mathbf{n}}^{\beta}\label{udu}
\\
dU_{x\mathbf{n}}U^{\dagger}_{x\mathbf{n}}
=+i\frac{e^{i\bar{V}_{x\mathbf{n}}}-1}{i\bar{V}_{x\mathbf{n}}} dv_{x\mathbf{n}}
=+iT^{\alpha} 
\left( \frac{e^{i\bar{V}_{x\mathbf{n}}}-1}{i\bar{V}_{x\mathbf{n}}} 
\right)_{\alpha\beta} dv_{x\mathbf{n}}^{\beta}\label{duu}
\end{align}
where $\bar{V}_{x\mathbf{n}}$ is defined by the adjoint representation 
of $U(2)$ gauge group given as 
\begin{equation}
\bar{V}_{x\mathbf{n}}^{bc}\equiv -i\epsilon^{abc}v^a_{x\mathbf{n}},
\quad (a,b,c= 1,2,3)
\end{equation}
This derivation is the same as described in Ref.~\cite{Kawai-Nakayama-Seo}.
Substituting 
Eqs.~(\ref{udu}) and (\ref{duu}) into Eq.~(\ref{norm}),
we obtain the explicit form of the metric 
$g_{\alpha\beta\mathbf{n}}
=g_{\alpha\beta\mathbf{n}}^{(x)}+g_{\alpha\beta\mathbf{n}}^{(y)}$, 
where 
\begin{align}
g_{\alpha\beta\mathbf{n}}^{(x)}&=\frac{1}{2}
\left(
\begin{array}{@{\,}cc@{\,}}
\mathbf{1} & \frac{1}{2}(F^{(2)}_{\alpha\beta}+F^{(3)}_{\alpha\beta})_{\mathbf{n}}\\
\frac{1}{2}(F^{(2)T}_{\alpha\beta}+F^{(3)T}_{\alpha\beta})_{\mathbf{n}} 
& \frac{1}{2}(H_{\alpha\beta}+H^T_{\alpha\beta})_{\mathbf{n}}
\end{array}\right),\\
F^{(2)}_{\alpha\beta\mathbf{n}}ds^{\alpha}_{\mathbf{n}}dv^{\beta}_{\mathbf{n}}
&=Tr(T^{\alpha}T^{\gamma}T^{\delta}) 
(-i\left( \frac{e^{i\bar{V}_{x\mathbf{n}}}-1}
{i\bar{V}_{x\mathbf{n}}} \right)_{\gamma\beta})
z_{x\mathbf{n}}^{\delta}ds^{\alpha}_{\mathbf{n}}
dv^{\beta}_{\mathbf{n}},\\ 
F^{(3)}_{\alpha\beta\mathbf{n}}ds^{\alpha}_{\mathbf{n}}dv^{\beta}_{\mathbf{n}}
&=Tr(T^{\delta}T^{\gamma}T^{\alpha}) z_{x\mathbf{n}}^{\delta}
(i\left( \frac{e^{i\bar{V}_{x\mathbf{n}}}-1}{i\bar{V}_{x\mathbf{n}}}
 \right)_{\gamma\beta} )ds^{\beta}_{x\mathbf{n}}dv^{\alpha}_{x\mathbf{n}},\\
H_{ab\mathbf{n}}dv^{\alpha}_{x\mathbf{n}}dv^{\beta}_{x\mathbf{n}}
&=Tr (T^{\epsilon}T^{\zeta}T^{\gamma}T^{\delta})
z_{x\mathbf{n}}^{\epsilon}z_{x\mathbf{n}}^{\zeta}
\left( \frac{e^{i\bar{V}_{x\mathbf{n}}}-1}{i\bar{V}_{x\mathbf{n}}} 
\right)_{\gamma\alpha} 
\left( \frac{e^{i\bar{V}_{x\mathbf{n}}}-1}{i\bar{V}_{x\mathbf{n}}} 
\right)_{\delta\beta}dv^{\alpha}_{\mathbf{n}}dv^{\beta}_{\mathbf{n}},
\end{align}
and similar expressions for $g_{\alpha\beta\mathbf{n}}^{(y)}$.
We note that the cross terms of $x$ and $y$ vanish.

The square root of the determinant of the metric is 
\begin{equation}
det(g)_{\mathbf{n}}=det(g^{(x)})_{\mathbf{n}}det(g^{(y)})_{\mathbf{n}},
\end{equation}
where 
\begin{align}
&\sqrt{det(g^x)_{\mathbf{n}}} 
=exp[\frac{1}{2}\sum_{\alpha,\beta} [log
\{\frac{1}{2}(H+H^T)_{\mathbf{n}}
-\frac{1}{4}(F^{(2)T}+F^{(3)T})_{\mathbf{n}}
\times(F^{(2)}+F^{(3)})_{\mathbf{n}}\}]_{\alpha\beta}],
\end{align}
and similar expression for $g^y_{\mathbf{n}}$.
When we ignore the gauge fields, $log \sqrt{det(g^x)_{\mathbf{n}}} $
reduces to 
\begin{align}
log \sqrt{det(g^x)_{\mathbf{n}}} =
log[\frac{1}{2}Tr (z_{x\mathbf{n}}^2\{T^{\alpha},T^{\alpha}\})
-\frac{1}{4}Tr(z_{x\mathbf{n}}[T^{\gamma},T^{\alpha}])Tr(z_{x\mathbf{n}}
[T^{\beta},T^{\gamma}])].
\end{align}
Expanding $log \sqrt{det(g^x)_{\mathbf{n}}}$ 
around the $s_x=s_y=0$ through second order, 
we obtain the effective action from the measure term 
\begin{align}
S_{meas}=&\sum_{\mathbf{n}}[-log[(1+\langle s \rangle)^2]M^2-(1+\langle s \rangle)^{-1}
Tr [(s_{x\mathbf{n}}+s_{y\mathbf{n}})\{T^{\alpha},T^{\alpha}\}]
\nonumber\\
&-(1+\langle s \rangle)^{-2}\frac{1}{2}
Tr [(s_{x\mathbf{n}}^2+s_{x\mathbf{n}}^2)\{T^{\alpha},T^{\alpha}\}]\nonumber\\
&+(1+\langle s \rangle)^{-2}\frac{1}{4}
Tr[(s_{x\mathbf{n}}+s_{y\mathbf{n}})[T^{\beta},T^{\alpha}]]
Tr[(s_{x\mathbf{n}}+s_{x\mathbf{n}})[T^{\alpha},T^{\beta}]]+ \cdots ]
\end{align}

\section{2-point amplitudes}\label{ampli}
The 2-point amplitudes corresponding to the Feynman diagrams for
scalar, ghost, gauge boson, and fermion loops as well as the measure term 
are given by
\\ 
{\footnotesize 
\begin{enumerate}
\begin{minipage}{.45\linewidth}
\item contribution from scalar 3-point vertex
\begin{align}
\frac{1}{N^2}\sum_{\mathbf{k}} &\frac{(1+\langle s \rangle)^2(\widehat{k_x}^4+\frac{3}{2}\bar{\mu}^2\widehat{\mathbf{k}}^2+\frac{9}{2}\bar{\mu}^4)}{[(1+\langle s \rangle)^2(\widehat{\mathbf{k}}^2+3\bar{\mu}^2)-\bar{\mu}^2]^2}2\delta^{\alpha_1,0}\delta^{\alpha_2,0}\nonumber\\
&+\frac{(1+\langle s \rangle)^2(\widehat{k_x}^4+\frac{3}{2}\bar{\mu}^2\widehat{\mathbf{p}}^2+\frac{9}{2}\bar{\mu}^4)}{[(1+\langle s \rangle)^2(\widehat{\mathbf{k}}^2+3\bar{\mu}^2)-\bar{\mu}^2]^2}2M\delta^{\alpha_1,\alpha_2}\nonumber
\end{align}
\end{minipage}
\begin{minipage}{.50\linewidth}
\includegraphics[width=6cm,clip]{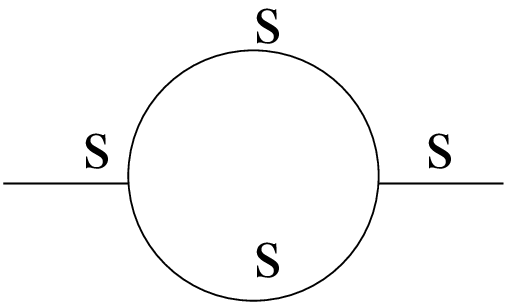}
\end{minipage}
\begin{minipage}{.50\linewidth}
\item contribution from scalar 4-point vertex
\begin{align}
\frac{1}{N^2}\sum_{\mathbf{k}} 
&\frac{\frac{3}{8}\widehat{\mathbf{k}}^2+1+\frac{3}{4}\bar{\mu}^2}
{(1+\langle s \rangle)^2(\widehat{\mathbf{k}}^2+3\bar{\mu}^2)-\bar{\mu}^2}
2\delta^{\alpha_1,0}\delta^{\alpha_2,0}\nonumber\\
&+\frac{-\frac{1}{8}\widehat{\mathbf{k}}^2-1+\frac{3}{4}\bar{\mu}^2}
{(1+\langle s \rangle)^2(\widehat{\mathbf{k}}^2+3\bar{\mu}^2)-\bar{\mu}^2}
2M\delta^{\alpha_1,\alpha_2}\nonumber
\end{align}
\end{minipage}
\begin{minipage}{.50\linewidth}
\includegraphics[width=6cm,clip]{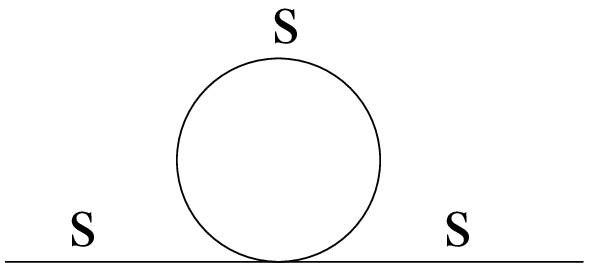}
\end{minipage}

\begin{minipage}{.50\linewidth}
\item contribution from scalar 2 gauge 1 vertex
\begin{align}
\frac{1}{N^2}\sum_{\mathbf{k}} 
&\frac{-\frac{1}{8}\widehat{\mathbf{k}}^2-1}
{(1+\langle s \rangle)^2(\widehat{\mathbf{k}}^2+3\bar{\mu}^2)-\bar{\mu}^2}
2\delta^{\alpha_1,0}\delta^{\alpha_2,0}\nonumber\\
+&\frac{\frac{1}{8}\widehat{\mathbf{k}}^2+1}
{(1+\langle s \rangle)^2(\widehat{\mathbf{k}}^2+3\bar{\mu}^2)-\bar{\mu}^2}
2M\delta^{\alpha_1,\alpha_2}\nonumber
\end{align}
\end{minipage}
\begin{minipage}{.50\linewidth}
\includegraphics[width=6cm,clip]{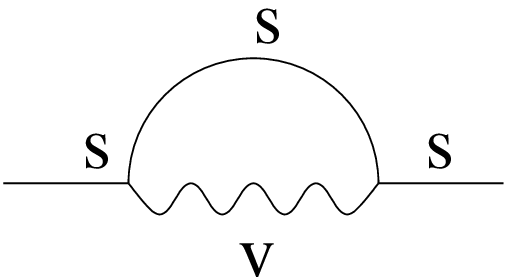}
\end{minipage}

\begin{minipage}{.50\linewidth}
\item contribution from ghost loop
\begin{align}
\frac{1}{N^2}\sum_{\mathbf{k}} 
&\frac{-\frac{1}{4}\widehat{k_x}^4}
{(1+\langle s \rangle)^2(\widehat{\mathbf{k}}^2)^2}2
\delta^{\alpha_1,0}\delta^{\alpha_2,0}\nonumber\\
&+\frac{-\frac{1}{4}\widehat{k_x}^4}
{(1+\langle s \rangle)^2(\widehat{\mathbf{k}}^2)^2}2M
\delta^{\alpha_1,\alpha_2}\nonumber
\end{align}
\end{minipage}
\begin{minipage}{.50\linewidth}
\includegraphics[width=6cm,clip]{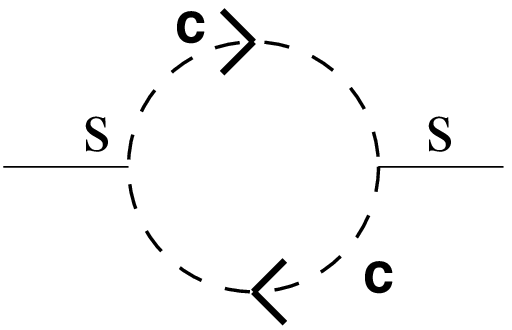}
\end{minipage}

\begin{minipage}{.50\linewidth}
\item contribution from gauge2 scalar2 vertex
\begin{align}
\frac{1}{N^2}\sum_{\mathbf{k}} &\frac{2-\frac{1}{8}\widehat{\mathbf{k}}^2}
{(1+\langle s \rangle)^2(\widehat{\mathbf{k}}^2)}
2\delta^{\alpha_1,0}\delta^{\alpha_2,0}\nonumber\\
&+\frac{-2-\frac{1}{8}\widehat{\mathbf{k}}^2}
{(1+\langle s \rangle)^2(\widehat{\mathbf{k}}^2)}
2M\delta^{\alpha_1,\alpha_2}\nonumber
\end{align}
\end{minipage}
\begin{minipage}{.50\linewidth}
\includegraphics[width=6cm,clip]{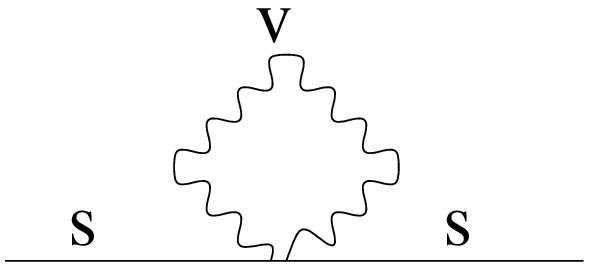}
\end{minipage}

\begin{minipage}{.50\linewidth}
\item contribution from gauge2 scalar1 vertex
\begin{align}
\frac{1}{N^2}\sum_{\mathbf{k}} 
&\frac{\frac{3}{4}(\widehat{\mathbf{k}}^2)^2
-\frac{1}{4}\widehat{k_x}^2\widehat{k_y}^2}
{(1+\langle s \rangle)^2(\widehat{\mathbf{k}}^2)^2}
2\delta^{\alpha_1,0}\delta^{\alpha_2,0}\nonumber\\
&+\frac{\frac{3}{4}(\widehat{\mathbf{k}}^2)^2
-\frac{1}{4}\widehat{k_x}^2\widehat{k_y}^2}
{(1+\langle s \rangle)^2(\widehat{\mathbf{k}}^2)^2}
2M\delta^{\alpha_1,\alpha_2}\nonumber
\end{align}
\end{minipage}
\begin{minipage}{.50\linewidth}
\includegraphics[width=6cm,clip]{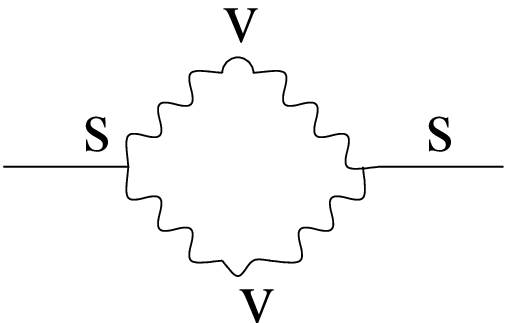}
\end{minipage}

\begin{minipage}{.50\linewidth}
\item contribution from measure term 
\begin{align}
\frac{1}{N^2}\sum_{\mathbf{k}} 
&\frac{-\frac{1}{2}(\widehat{\mathbf{k}}^2)^2}
{(1+\langle s \rangle)^2(\widehat{\mathbf{k}}^2)^2}
2\delta^{\alpha_1,0}\delta^{\alpha_2,0}\nonumber\\
&+\frac{-\frac{1}{2}(\widehat{\mathbf{k}}^2)^2}
{(1+\langle s \rangle)^2(\widehat{\mathbf{k}}^2)^2}
2M\delta^{\alpha_1,\alpha_2}\nonumber
\end{align}
\end{minipage}
\begin{minipage}{.50\linewidth}
\includegraphics[width=6cm,clip]{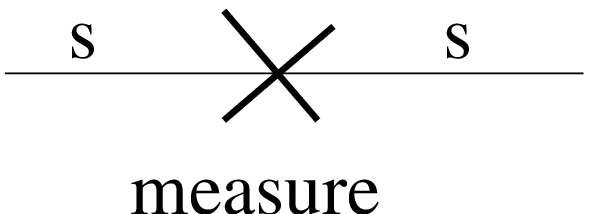}
\end{minipage}

\begin{minipage}{.50\linewidth}
\item contribution from fermion loop
\begin{align}
\frac{1}{N^2}\sum_{\mathbf{k}} 
2\delta^{\alpha_1,0}\delta^{\alpha_2,0} F_{u1}\nonumber\\
2M\delta^{\alpha_1,\alpha_2} F_{su2} \nonumber
\end{align}
\end{minipage}
\begin{minipage}{.50\linewidth}
\includegraphics[width=6cm,clip]{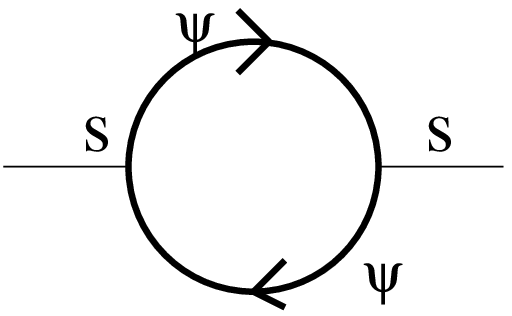}
\end{minipage}
\end{enumerate}}
\begin{align}
&F_{u1}=\frac{(F^{(1)}_{u1}+F^{(2)}_{u1})}
{(1+\langle s \rangle )^2[\widehat{\mathbf{k}}^2(1+\bar{\mu_f})+2\bar{\mu_F}^2]^2}
\nonumber\\
&F^{(1)}_{u1}=-(2\widehat{k_x}^2(1+\bar{\mu_F})+\bar{\mu_F}^2)(1+\bar{\mu_F})
\quad 
F^{(2)}_{u1}=-2[(1+\bar{\mu_F})^2\widehat{k_y}^2-\bar{\mu_F}^2](1+\bar{\mu_F})
\cos k_x \nonumber\\
&F_{su2}=\frac{(F^{(1)}_{su}+F^{(2)}_{su}+F^{(3)}_{su}+F^{(4)}_{su})}
{(1+\langle s \rangle )^2[\widehat{\mathbf{k}}^2(1+\bar{\mu_f})+2\bar{\mu_F}^2]^2}\\
&F^{(1)}_{su}=\widehat{k_x}^2 \cos (k_x)[(1+\bar{\mu_F})^2+1] \quad
F^{(2)}_{su}=-\bar{\mu_F}(1+\bar{\mu_F})^2\widehat{k_x}^2
+\bar{\mu_F}\widehat{k_x}\widehat{3k_x}\nonumber\\
&F^{(3)}_{su}=\bar{\mu_F}^2(1+\bar{\mu_F})^2+\bar{\mu_F}\cos (2k_x) \quad 
F^{(4)}_{su}= [(1+\bar{\mu_F})^2\widehat{k_y}^2+\bar{\mu_F}^2]
[(1+\bar{\mu_F})^2+1]
\end{align}
\newpage

\end{document}